\documentclass{aa}
\usepackage{graphics,psfig}
\begin{document}


\title{Hard X--ray properties of blazars}

\author{}
\institute{}

\author{D. Donato\inst{1} 
\and G. Ghisellini\inst{1} 
\and G. Tagliaferri\inst{1}
\and G. Fossati\inst{2} }

\offprints{G. Ghisellini
}

\institute{Osservatorio Astronomico di Brera, Via Bianchi 46, I-23807 Merate,
Italy
\and Center for Astrophysics and Space Sciences, University of California
     at San Diego, 9500 Gilman Drive, La Jolla, CA 92093-0424, USA
}

\date{Received date / accepted date}

\titlerunning{Hard X--ray properties of blazars}
\authorrunning{D. Donato et al.\ }

\abstract{
We have considered all blazars observed in the X--ray band and
for which the slope of the X--ray spectrum is available.
We have collected 421 spectra of 268 blazars, including 
12 archival unpublished ASCA spectra of 7 blazars whose analysis
is presented here.
The X--ray spectra of blazars show trends as a function
of their power, confirming that the blazar overall energy
distribution can be parameterized on the basis of one parameter
only, i.e. the bolometric luminosity.
This is confirmed by the relatively new hard (2--10 keV)
X--ray data. 
Our results confirm the idea that in low power objects the
X--ray emission mechanism is the synchrotron process, 
dominating both the soft and the hard X--ray emissions.
Low energy peaked BL Lac objects are intermediate, often showing
harder spectra in the hard X--ray band, suggesting that the synchrotron
process dominates in the soft band, with the inverse Compton process
dominating at high energies.
The most powerful objects have X--ray spectra that are flat both in the
soft and in the hard band, consistent with a dominating inverse Compton 
component.
}
\maketitle

\section{Introduction}

Thanks to $\gamma$--ray observations of EGRET, on board CGRO, we now
know the overall spectral energy distribution (SED) of blazars.
They are characterized, in $\nu$--$\nu F(\nu)$ plots, by two
broad peaks. 
It is believed that the first, located in the IR--soft X--ray band,
is due to synchrotron emission, while the second is due to the
inverse Compton process by the same electrons producing the
synchrotron part of the spectrum (Maraschi, Ghisellini \& Celotti 1992; 
Sikora, Begelman \& Rees 1994; but see Mannheim 1993
for a different interpretation).
The 0.1--10 keV emission of blazars is therefore located in the
minimum between the two peaks, where both processes (synchrotron
and inverse Compton) can contribute.
Observations in this band are therefore useful to characterize 
the relative importance of both processes.
This can constrain models, allowing a better determination
of the location of both peaks.
Usually, a steep power law energy distribution in the X--ray band
(with a spectral energy index $\alpha>1$, with $F \propto \nu^{-\alpha}$)
is due to the tail of the synchrotron spectrum,
while $\alpha<1$ flags the dominance of the inverse Compton spectrum.
There are exceptions to this rule, such as HBL (High energy peaked BL Lacs)
in a flaring state, which show a synchrotron spectrum peaking above 10 keV.
One dramatic example of this behavior is Mkn 501, whose synchrotron
peak energy, usually located below/at $\simeq$1 keV, shifted to 100 keV or
more during its flare in April 1997 (Pian et al. 1998).
In these cases the X--ray spectrum, usually steep during
quiescence, becomes much flatter during flares.
 
Fossati et al. (1998) (F98 hereafter) have shown that blazars form a sequence,
with their SED changing in a continuous way as their bolometric
power changes:
low luminosity objects (HBL) have the synchrotron peak in the UV--soft X--ray 
band, and the inverse Compton peak between the GeV and the TeV band.
The two components have approximately the same power.
As the bolometric luminosity increases, both peaks shift to lower
frequencies, and the Compton peak becomes increasingly dominant.
This trend offers the opportunity to unify in a single scheme the many 
flavors of existing blazars, and calls for a physical explanation
(see e.g. Ghisellini et al., 1998).

To check the reliability of this trend we have collected data
for all blazars having available spectral information
in the X--ray band.
For the soft band [0.1--2 keV] most of the results come from {\it ROSAT},
while for the 2--10 keV band the results are gathered from the EXOSAT, ASCA
and {\it Beppo}SAX satellites. 

Besides the data already published, we searched for unpublished
data in the ASCA public archive, finding 12 observations of 7 sources.
Results of the analysis of these data are presented here.
We then add these sources to our sample.

The entire sample forms the largest database in the 
X--ray range: 421 spectra of 268 blazars.
The X--ray data have been complemented by additional information regarding
the redshift (when available), the radio flux at 5 GHz and the optical flux
(in the $V$ band).

The paper is organized as follows; section~\ref{sec:ASCA} is devoted to the
analysis of the 12 ASCA spectra, while in Section~\ref{sec:catalogue} we
present the entire set of data.
In Section~\ref{sec:results} we compare the results in the soft and the
hard X--ray bands and check for correlations with other spectral parameters,
such as the broad band spectral indices connecting
the radio with the optical fluxes, the optical with the X--ray fluxes,
and the radio with the X--ray fluxes.
In Section~\ref{sec:average_sed} we discuss our findings in the framework
of the scenario proposed by Fossati et al. (1998), suggesting an
improvement connected to a possible physical difference between low and
high power sources.

\section{Analysis of ASCA data}
\label{sec:ASCA}

We searched the ASCA public archive at HEASARC, finding 12 observations of
7 blazars that have not been analyzed and published before: 0405--123 and
PKS 0420--014 (classified as Flat Spectrum Radio Quasars, FSRQs); B2
1308+326 and 1807+698 (classified as Low Peaked BL Lac objects, LBLs); 1ES
1028+511, 1553+511 and 1ES 2344+514 (classified as High Peaked BL Lac
objects, HBLs). 

1ES 1028+511 and 1ES 2344+514 have 3 separate observations each, while 
B2 1308+326 was observed twice. 
The search for unpublished observations is updated to November 1999. 
During the preparation of this work the ``Tartarus" data base
became available\footnote{at {\tt http://tartarus.gsfc.nasa.gov/}}, 
presenting results of an automatic spectral and temporal analysis for the
AGNs observed by ASCA.

\subsection{Data Reduction and Spectral Analysis}

We extracted the spectra of all sources from the files produced
by the Revision--2 data release
and included data transmitted in all 3 modes (High, Medium and Low) to 
increase the signal/noise ratio. 
The event files are obtained from all four
instruments on board ASCA: the Solid--State Imaging Spectrometers (SIS0 and 
SIS1) and the Gas Imaging Spectrometers (GIS2 and GIS3). For the SIS we used 
the event files converted into BRIGHT mode. 
For a description of the ASCA observatory see e.g. Tanaka et al. (1994).

To screen the SIS and GIS data we follow the criteria given
in the {\it ABC ASCA reduction guide}, rejecting the data taken 
during the passage of the South Atlantic Anomaly, or with
geomagnetic cutoff rigidity lower than 8 GeV/c, or with angles 
between the targets and the day/night terminator smaller than 20$^\circ$ 
or for Elevation angles smaller than 5$^\circ$.

The source spectra were extracted from circular regions centered 
on the sources, with radii of 6 arcmin for the GIS 
and 4 arcmin for the SIS0, while for the SIS1 the source is
normally nearer to the detector border and we had to use 
a smaller radius of $\sim 3.3$ arcmin. 
For the GIS we extracted the background in circular 
regions with the same dimensions used for the sources but centered on a 
symmetric point with respect to the optical axis, where the contribution 
of the source to the counts was negligible. 
For the SIS, instead, the background was extracted from
blank field files because the sources occupied a large area of the detector. 
On these blank fields we chose circular regions with the same radii 
and positions used for the sources.

For the GIS spectra we used the 1994 May response matrices, while 
for the SIS spectra we generated the matrices with the SISRMG program 
of the FTOOLS V3.6 package. The ARF files for both SIS and GIS
were derived with the ASCAARF V2.62 program. 
The GIS and SIS data were fitted in the channel ranges 
69--1020 and 15--510, corresponding to the
energy ranges 0.7--10 and 0.4--10 keV, respectively. 
The spectra were rebinned in order to have at least 25
counts in each new bin.

The ASCA spectra were fitted using XSPEC V10 with four models:
single or broken power law with free or fixed
(Galactic) absorption.
The cross section for photoelectric absorption is calculated following
Morrison \& McCammon (1983), while the Galactic column density in the
direction of the sources was estimated from the 21cm radio maps of neutral
hydrogen (Brinkmann \& Siebert, 1994; Danly et al., 1992; Dickey \& Lockman,
1990; Elvis et al., 1989; Lamer, Brunner \& Staubert, 1996; Lockman \& Savage 
1995; Murphy et al. 1996).
The data were fitted simultaneously from all the instruments with the same
model.  However, the normalizations were left as independent parameters 
for each data set to account for the cross--calibration uncertainties
between the four detectors, estimated to be of the order of 6\%.
The differences found between the various normalizations 
were always consistent with these uncertainties.


\begin{table*}
\begin{center}
\begin{tabular}{lllccclcc}
\hline
Source  &Obs. date$^a$  &$N_{\rm H}$      
&$\Gamma$ 
&$\chi_r^2/dof$ &$F_{[2-10]}^b$ &$F_{1keV}$\\
        &               &$10^{21}$cm$^{-2}$
&           &               &               &$\mu Jy$  \\
\hline
0405$-$123 &09/08/1998 &0.72$^{+0.51}_{-0.63}$ &1.76$^{+0.09}_{-0.10}$  &1.0/160   &5.2  &0.9   \\
0420$-$014 &31/08/1997 &0.90$^{+1.04}_{-0.81}$ &1.86$^{+0.19}_{-0.18}$  &0.8/73    &1.4  &0.3   \\
1028$+$511 &28/04/1995 &1.01$^{+0.23}_{-0.21}$ &2.53$^{+0.06}_{-0.05}$  &1.0/287   &6.2  &3.4   \\
           &29/04/1995 &1.33$^{+0.18}_{-0.17}$ &2.59$^{+0.05}_{-0.05}$  &0.9/310   &7.4  &4.5   \\
           &08/05/1995 &1.12$^{+0.12}_{-0.12}$ &2.49$^{+0.04}_{-0.04}$  &0.9/201   &7.8  &4.1   \\
1308$+$326 &10/06/1996 &1.97$^{+2.96}_{-1.97}$ &1.99$^{+0.47}_{-0.36}$  &1.5/29    &0.5  &0.1   \\
           &11/06/1996 &0.97$^{+1.01}_{-0.97}$ &1.74$^{+0.32}_{-0.23}$  &1.4/78    &0.6  &0.1   \\
1553$+$113 &16/08/1995 &1.30$^{+0.62}_{-0.61}$ &2.47$^{+0.19}_{-0.18}$  &1.3/294   &29.4 &14.9  \\
1807$+$698 &05/11/1996 &0.50$^{+0.28}_{-0.27}$ &1.75$^{+0.07}_{-0.06}$  &1.0/122   &3.1  &0.5   \\
2344$+$514 &10/01/1997 &2.71$^{+0.17}_{-0.17}$ &2.13$^{+0.03}_{-0.03}$  &1.1/216   &17.2 &5.3   \\
           &23/01/1997 &2.91$^{+0.34}_{-0.33}$ &2.39$^{+0.08}_{-0.06}$  &0.9/72    &10.4 &5.2   \\
           &10/12/1997 &2.93$^{+0.37}_{-0.35}$ &2.31$^{+0.07}_{-0.07}$  &1.2/202   &10.4 &4.2   \\
\hline
\end{tabular}
\caption{Best fits of the 12 observations.
a) day/month/year;
b) $10^{-12}$erg cm$^{-2}$s$^{-1}$ }
\end{center}
\label{tab:fits}
\end{table*}

\subsection{Results of the fits}

The results of the 12 spectral fits of the 7 blazars
observed by ASCA are reported in Table~1. 
The uncertainties for the spectral parameters are at the 90\% 
confidence errors for two parameters of interest ($\Delta \chi^2=$ 4.6). 
The unabsorbed integrated 2--10 keV and monochromatic 1 keV fluxes 
are obtained using only the SIS0 data. 
%
As the observations presented in this paper were all taken after 1994,
they are most likely affected by the so-called "excess $N_{\rm H}$"
problem. This is due to a degradation of the SIS efficiency below 1
keV which can give incorrect results for the column density and/or
other parameters. As suggested in the ASCA Web site, to avoid the 
calibration uncertainties we considered
only the SIS data above 1 keV in the SIS$+$GIS combined fit.
Of the four models considered, 
according to the F--test, the one that better represents the data 
in all twelve cases is the single power law with free $N_{\rm H}$.
In some cases (detailed below) we obtain a value for the absorbing
column greater (by a factor 3--10) than the Galactic value.
Note that the results obtained by the automatic analysis presented 
in the ``Tartarus" database are in excellent agreement with ours
(not surprisingly, since the same model is adopted).
What is somewhat surprising is that the broken power law model (either with
free or fixed $N_{\rm H}$) did not significantly improve the fits.
This may be indicative of true extra--absorption or a spectral behavior
more complex than those here examined (one possibility being a gradual but
continuous steepening of the spectrum).

In the following we will compare the spectral properties of blazars
in the soft and hard X--ray bands considering only the single power law model.
To this end the results of this model allow a more uniform comparison.

The results of spectral fit for the 7 sources are discussed below, grouped 
in the three subclasses (FSRQ, LBL and HBL).

\subsubsection{FSRQ}

For 0405--123, the best fit gives a flat photon spectral index
$\Gamma = 1.76 \pm 0.1$, indicating the dominance of the 
inverse Compton component. 
The derived $N_{\rm H}$ value is consistent with the Galactic value, 
$N_{\rm H}^{Gal} = 0.37 \times 10^{21}$ cm$^{-2}$ (Danly et al. 1992).

Similar results are obtained for PKS\,0420--014, with a photon
spectral index $\Gamma = 1.86 \pm 0.19$. 
Also in this case the $N_{\rm H}$ value is consistent with the Galactic one 
($N_{\rm H}^{Gal} = 0.94 \times 10^{21}$ cm$^{-2}$, Elvis et al. 1989).

\subsubsection{LBL}

For B2\,1308+326 there are two observations on two consecutive days.
The spectra are quite noisy and also the reduced $\chi_r^2$ are not
very good. In both cases the best fits are obtained with 
a spectral index $\Gamma \simeq$1.75--2.0.
Due to the large error bars (see Table~\ref{tab:fits}),
the derived absorption column density
can be consistent with the Galactic one ($N_{\rm H}^{Gal} = 0.11
\times 10^{21}$ cm$^{-2}$, Lockman \& Savage 1995).

For 1807+698 we obtained $\Gamma = 1.75\pm0.07$ and a value of $N_{\rm H}$
consistent with the Galactic one ($N_{\rm H}^{Gal} = 0.44 \times 10^{21}$
cm$^{-2}$, Murphy et al. 1996).

\subsubsection{HBL}

1ES\,1028+511 has been observed three times over a time span of two weeks.
The source did not significantly vary in flux nor in shape
($\Delta\Gamma \simeq 0.12$). 
In all cases the best fits are obtained for a value of $N_{\rm H}$ 
a factor 10 larger than the Galactic one ($N_{\rm H}^{Gal} = 0.12 
\times 10^{21}$ cm$^{-2}$; Lamer, Brunner \& Staubert, 1996).

For 1553+113 the best fit is obtained with a value of $N_{\rm H}$ $\sim 3$
times larger than the Galactic one ($N_{\rm H}^{Gal} = 0.37 \times 10^{21}$
cm$^{-2}$; Brinkmann \& Siebert, 1994).

The source 1ES\,2344+514 has been observed three times, twice in Jan 1997
and one in Dec 1997.
Both the fluxes and the spectral indices show some variability. 
The best fit value for the $N_{\rm H}$ is about 1.5 times the Galactic value
($N_{\rm H}^{Gal} = 1.67 \times 10^{21}$ cm$^{-2}$
Dickey \& Lockman, 1990).

\section{The catalogue}
\label{sec:catalogue}

\subsection{Starting samples}

Our purpose is to have the most complete ensemble of spectral information
(fluxes and spectral indices) in the X--ray band, from 0.1 to 10 keV, of all 
known blazars. 
We therefore considered all blazars detected in the X--ray band,
for which also a measure of the X--ray spectral index is available.
We collected the data obtained by five X--ray satellites: {\it Einstein},
EXOSAT, {\it ROSAT}, ASCA and {\it Beppo}SAX (see Table \ref{tab:sat_obs}).

The first step was to recognize if a source belongs to the blazar class, 
and to which subclass (i.e. if a source is a FSRQ or an HBL or an LBL).
We used several published lists of blazars and other publications 
describing single recognized sources. 
We considered the Slew Survey Sample (Elvis et al. 1992, Perlman et al.
1996), the 2 Jy sample of Wall \& Peacock (1985), and the 1 Jy BL Lac
sample (Stickel et al. 1991).
In addition, we used the lists taken from the works of Bade et al. (1994), 
Bade et al. (1998), Brinkmann et al. (1994), Brinkmann et al. (1997), Cappi
et al. (1997), Comastri et al. (1997), Ghisellini et al. (1993), Lamer et
al. (1996), Laurent--Muehleisen et al. (1999), Sambruna et al. (1997),
Wolter et al. (1998) and Worrall et al. (1990). 
We also checked the NASA Extragalactic Database (NED) for other objects
classified as BL Lacs/blazars, or that could be classified as such. 

The total number of considered blazars is 268.
Of these, 227 have been observed by {\it ROSAT} and 88 have spectral
information in the 2--10 keV band [of these latter sources, 77 have both
soft ({\it ROSAT}) and hard X--ray data].
The details about the breakdown of source among HBL/LBL/FSRQ, and of the
data among different X--ray telescopes is reported in Table~\ref{tab:sat_obs}.
The data obtained with {\it Einstein} 
have large errors associated and for almost all sources
better {\it ROSAT} data were available. 
For these reasons, the Einstein data
are not used to derive any of the results (or figures) of this paper.

Some sources have been observed many times either by the same and/or by
different satellites. 
For these sources, we chose the observation with the best $\chi_r^2$ in 
the analysis
{\footnote{ 
We have verified that this choice does not introduce a bias towards 
particularly faint states of the sources (for which lower statistics 
could yield larger error bars and better $\chi^2$)
nor towards particularly high states (for which a better statistics
could yield lower values of the reduced $\chi^2$ due to the increased
number of degrees of freedom).}}.


For the most ``famous" sources, like 3C 273, Mkn 421, Mkn 501, PKS 2155--304,
we do not include the results of all the observations made by all satellites,
but we have only listed few representative data (those with the best
$\chi_r^2$) for each of these sources (typically, one spectral datum for
each observing satellite).

Of course, the resulting catalogue is not a complete sample.
Nevertheless, it is the largest database of its kind, and we think it is
representative of the entire blazar class.
The large number of sources in each sub--category of blazars guarantees 
a meaningful comparison between their
X--ray properties, and their relation with the fluxes in other bands.
%

\begin{table}
\begin{center}
\begin{tabular}{|l|c|ccc|}
\hline
        &$N^o$oss.& HBL & LBL & FSRQ \\
\hline
ASCA          & \phantom{2}52     & 14 &  9 & 24 \\
EXOSAT        & \phantom{2}33     & 16 &  7 & 10 \\
BeppoSAX      & \phantom{2}47     & 29 &  9 &  8 \\
{\it ROSAT}   &           227     &129 & 54 & 44 \\
EINSTEIN      & \phantom{2}62     &  7 & 23 & 32 \\
\hline
TOTAL         &           421     &136 & 63 & 69 \\
\hline
\end{tabular}
\end{center}
\caption{Number of observations obtained from various satellites
and number of observed blazars (divided into different subclasses). 
For the total number of sources we have excluded multiple
observations by different satellites of the same source.}
\label{tab:sat_obs} 
\end{table}

\subsection{Format of the catalogue}
\label{sec:format}

Data are presented in Table~\ref{tab:listone} with the following format.
For each source, Table~\ref{tab:listone} gives the IAU name, the redshift,
the fluxes in the radio band (5 GHz), optical ($V$ band) and X--ray (1 keV)
and the X--ray photon spectral index.
In the last columns of Table~\ref{tab:listone} we also indicate to which
subclass the blazar belongs to (1 for HBL, 2 for LBL and 3 for FSRQs) and
the observing satellite (RO={\it ROSAT}; AS=ASCA; EI={\it Einstein};
EX=EXOSAT; SA={\it Beppo}SAX).

For the radio fluxes we calculated the averaged value when 
there was more than one observation; the optical fluxes reported in the NED 
database are calculated using the indicated magnitude dereddened with the 
galactic extinction $A_B$ as reported by the NED database. 
When in the literature we found only the 0.1--2.4 keV and/or the 2--10 keV 
integrated fluxes, we derived the monochromatic ones at 1 keV using the
corresponding X--ray spectral index.
All fluxes presented in Table~\ref{tab:listone} are not K--corrected.

To compute the luminosities, we used $H_0$=50 km s$^{-1}$ Mpc$^{-1}$ and
$q_0$=0.5, and for the K--correction we assumed a radio spectral index
$\alpha=0$ for all sources; an optical spectral index 
$\alpha = 0.5$ for HBL and $\alpha =1$ for the rest of the sources;
for the X--ray data we used the listed X--ray spectral index.
Also the broad band spectral indices have been K--corrected.

The K--correction for sources with unknown redshift was computed using the
average redshift appropriate for each sub--class (i.e. $\langle z
\rangle_{HBL}= 0.249$, $\langle z \rangle_{LBL}= 0.457$ and $\langle z
\rangle_{FSRQ}= 1.265$).

\section{Results}
\label{sec:results}

Theoretical models that explain the nature of the observed behaviors of 
blazars predict a continuity between the various subclasses. 
To check if this is true for the objects belonging to our catalogue, 
we computed the distributions of redshifts, X--ray and broad band spectral
indices and luminosities. 

Moreover, another goal of this work is to compare the blazar
characteristics observed in the soft and in the hard X--ray band. 
We divided our sources into two groups, one with the data obtained 
with {\it ROSAT} (0.1--2.4 keV) and another one with the data obtained with 
EXOSAT, ASCA, and {\it Beppo}SAX (2--10 keV). 
As anticipated (see Table~\ref{tab:sat_obs}) the first group of sources
contains 227 objects, while the second one contains 88 sources (38 HBL, 19
LBL and 31 FSRQ). 
In addition, since the two sources 2344+514 and 1652+398 are very variable
and we have data both for a quiescent and a flaring state, in the latter
group we put the data of two observations (one for the high and one for the
low state) for each of them. 

\begin{figure}
\psfig{figure=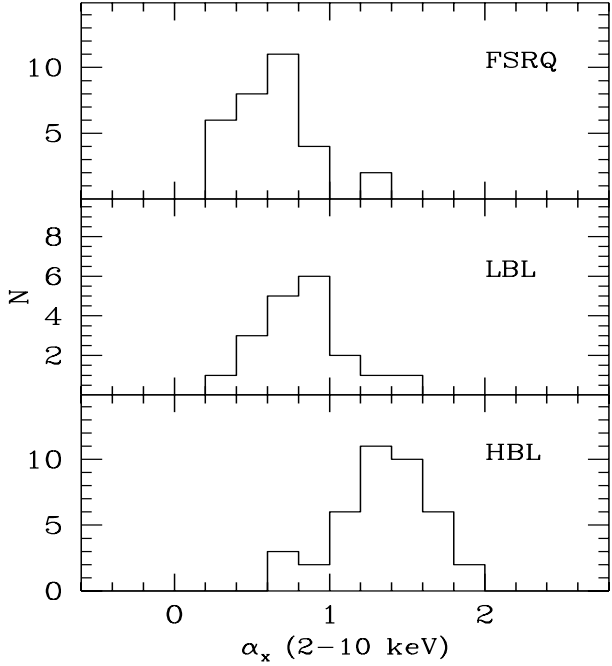,width=11.5cm,height=9.5cm}
\vspace{-0.3cm}
\caption{Distribution of the energy spectral index $\alpha_x$ in the 
2--10 keV hard X--ray energy band. Note the difference between 
the HBL and the other two subclasses of blazars.
The KS test gives a probability $P=8\times 10^{-6}$ that the HBL and the LBL
values are drawn from the same distribution
($P=2\times 10^{-12}$ for HBL--FSRQ and $P=0.03$ for LBL--FSRQ).
\label{fig:alpha_hard} }
\end{figure}
\begin{figure}
\psfig{figure=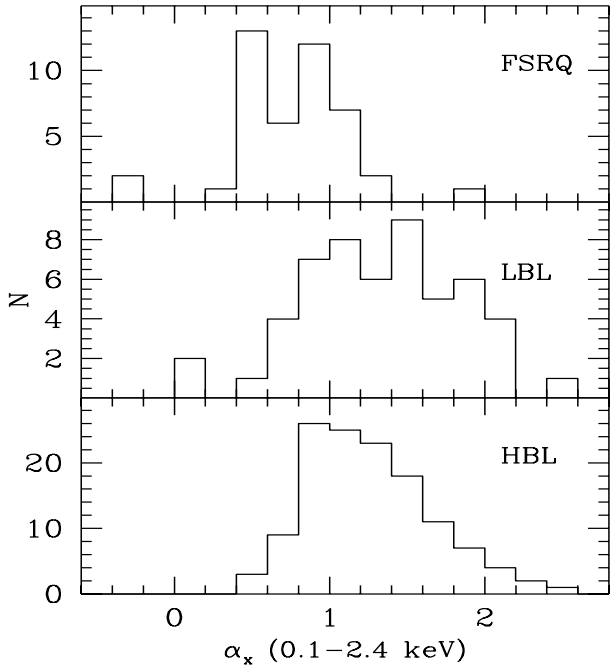,width=11.5cm,height=9.5cm}
\vspace{-0.3cm}
\caption{Distribution of the energy spectral index $\alpha_x$ in the 
0.1--2.4 keV soft X--ray energy band. 
In this case LBL are more similar to HBL than to FSRQ.
KS test results: 
$P=0.20$ for HBL--LBL; 
$10^{-8}$ for HBL--FSRQ; 
$5\times 10^{-8}$ for LBL--FSRQ.
\label{fig:alpha_soft} }
\end{figure}

\begin{figure}
\psfig{figure=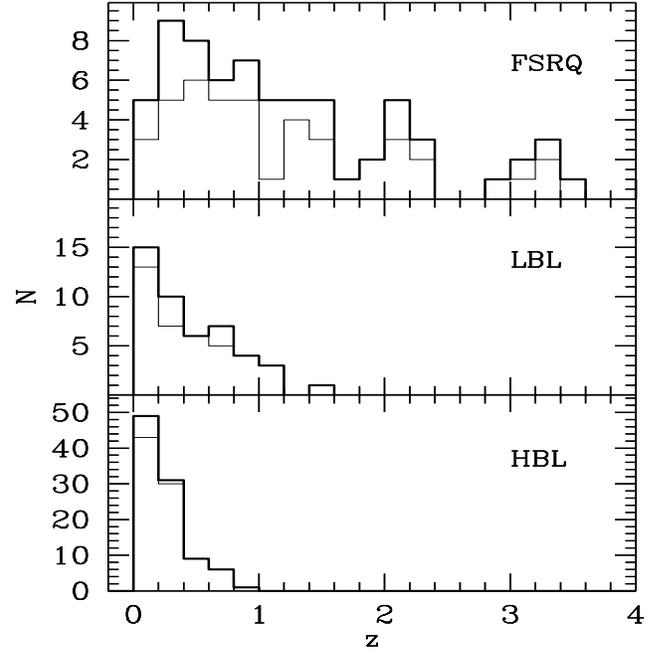,width=11.5cm,height=9.5cm}
\vspace{-0.3cm}
\caption{Redshift distribution for the three subclasses.
Thick solid lines refer to the entire sample, while
thin solid lines refer to sources with only {\it ROSAT} data.
KS test results (for the entire sample): 
$P=2 \times 10^{-3}$ for HBL--LBL; 
$9\times 10^{-17}$ for HBL--FSRQ; 
$5\times 10^{-5}$  for LBL--FSRQ.
\label{fig:z} }
\end{figure}
\begin{figure}
\psfig{figure=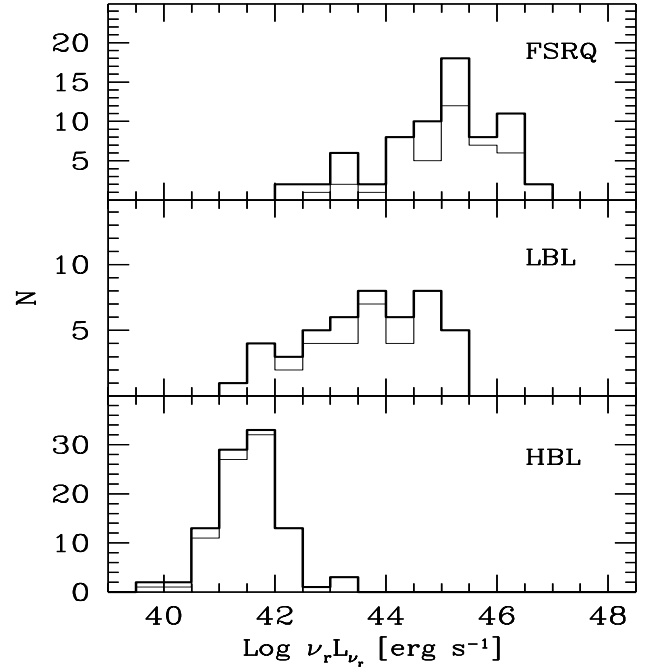,width=11.5cm,height=9.5cm}
\vspace{-0.3cm}
\caption{Distribution of the radio luminosity for the three subclasses.
Thick solid lines refer to the entire sample, while
thin solid lines refer to sources with only {\it ROSAT} data.
KS test results (for the entire sample): 
$7 \times 10^{-19}$ for HBL--LBL; 
$7 \times 10^{-33}$ for HBL--FSRQ; 
$10^{-6}$ for LBL--FSRQ.
\label{fig:lr}}
\end{figure}
\begin{figure}
\psfig{figure=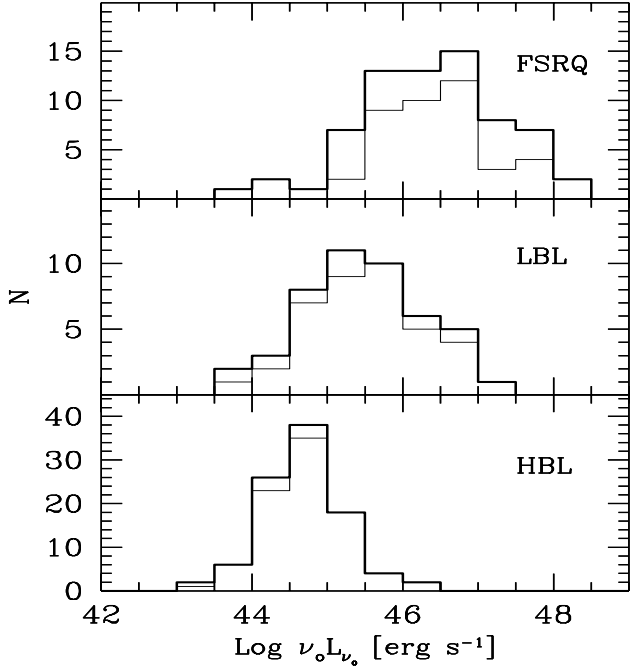,width=11.5cm,height=9.5cm}
\vspace{-0.3cm}
\caption{Distribution of the optical luminosity for the three subclasses.
Thick solid lines refer to the entire sample, while
thin solid lines refer to sources with only {\it ROSAT} data.
KS test results (for the entire sample): 
$P=2 \times 10^{-7}$ for HBL--LBL; 
$10^{-22}$ for HBL--FSRQ; 
$3 \times 10^{-5}$ for LBL--FSRQ.
\label{fig:lo} }
\end{figure}
\begin{figure}
\psfig{figure=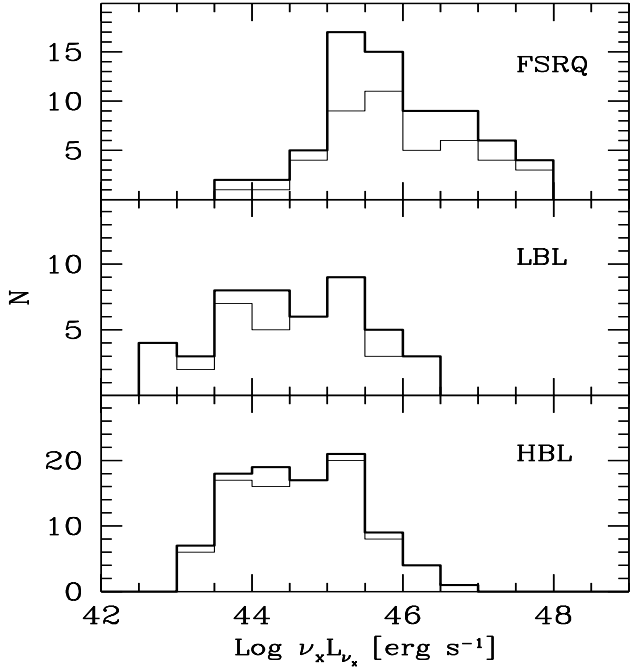,width=11.5cm,height=9.5cm}
\vspace{-0.3cm}
\caption{Distribution of the X--ray luminosity for the three subclasses.
Thick solid lines refer to the entire sample, while
thin solid lines refer to sources with only {\it ROSAT} data.
KS test results (for the entire sample): 
$P=0.6$ for HBL--LBL; 
$3 \times 10^{-13}$ for HBL--FSRQ ; 
$2 \times 10^{-8}$ for LBL--FSRQ.
\label{fig:lx} }
\end{figure}
\begin{figure}
\psfig{figure=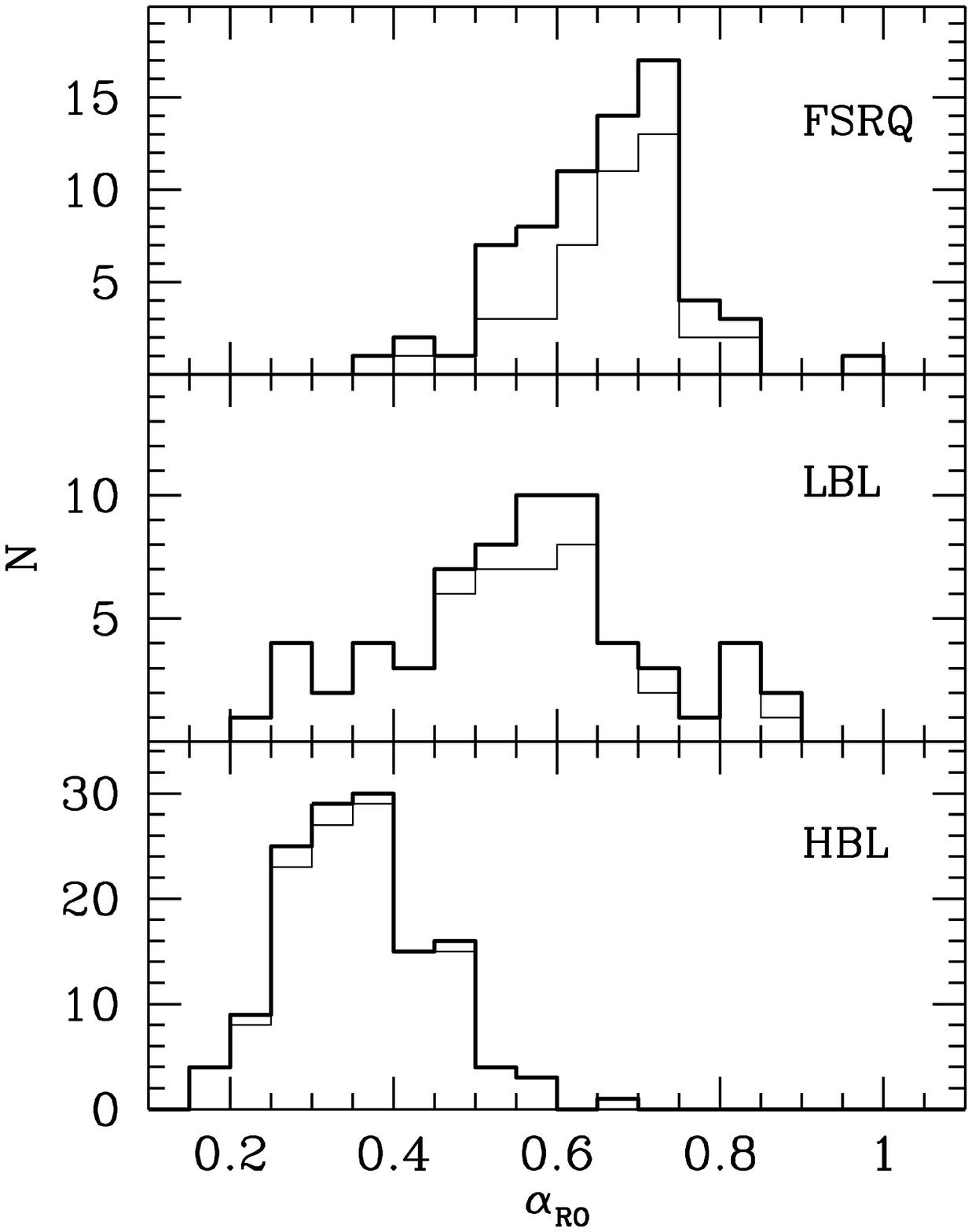,width=11.5cm,height=9.5cm}
\vspace{-0.3cm}
\caption{Distribution of the broad band radio--optical
spectral index for the three subclasses.
Fluxes have been K--corrected as explained in the text.
Thick solid lines refer to the entire sample, while
thin solid lines refer to sources with only {\it ROSAT} data.
KS test results (for the entire sample): 
$P=4 \times 10^{-18}$ for HBL--LBL; 
$2\times 10^{-35}$ for HBL--FSRQ; 
$4\times 10^{-4}$ for LBL--FSRQ.
\label{fig:alpha_ro}}
\end{figure}
\begin{figure}
\psfig{figure=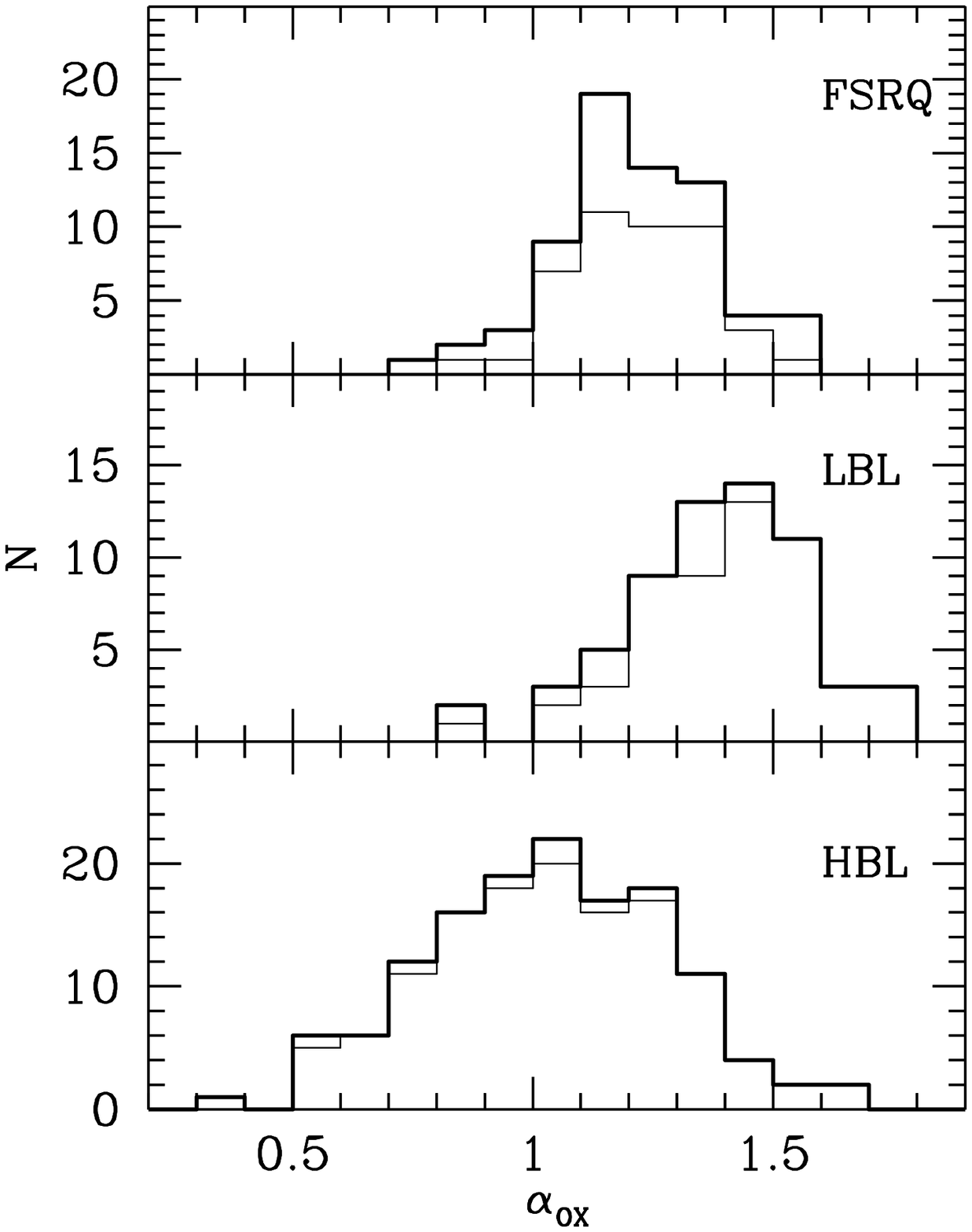,width=11.5cm,height=9.5cm}
\vspace{-0.3cm}
\caption{Distribution of the broad band optical--X--ray
spectral index for the three subclasses.
Fluxes have been K--corrected as explained in the text.
Thick solid lines refer to the entire sample, while
thin solid lines refer to sources with only {\it ROSAT} data.
KS test results (for the entire sample): 
$P=5 \times 10^{-17}$ for HBL--LBL; 
$10^{-5}$ for HBL--FSRQ; 
$10^{-8}$ for LBL--FSRQ.
\label{fig:alpha_ox} }
\end{figure}
\begin{figure}
\psfig{figure=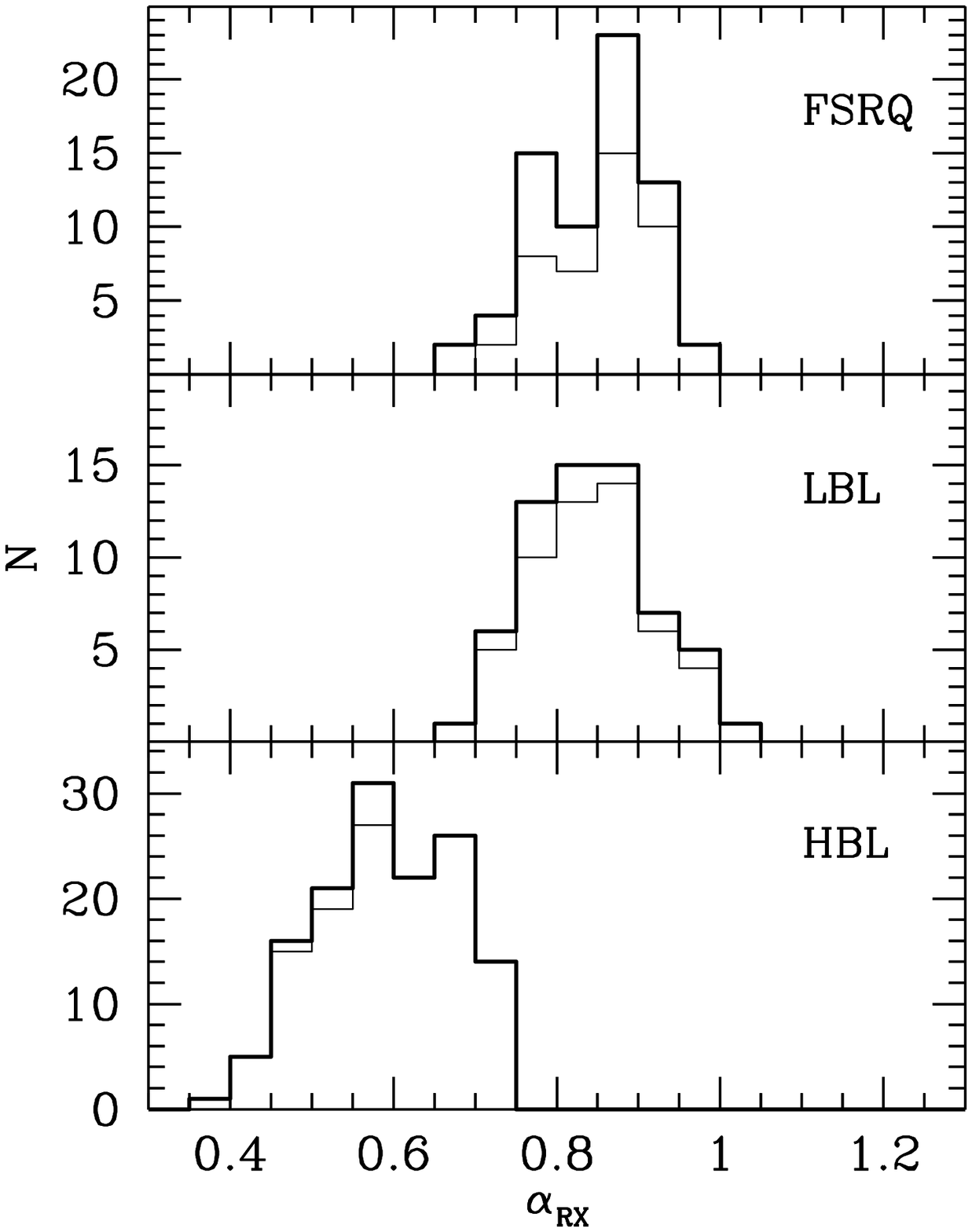,width=11.5cm,height=9.5cm}
\vspace{-0.3cm}
\caption{Distribution of the broad band radio--X--ray
spectral index for the three subclasses.
Fluxes have been K--corrected as explained in the text.
Thick solid lines refer to the entire sample, while
thin solid lines refer to sources with only {\it ROSAT} data.
KS test results (for the entire sample): 
$P=8 \times 10^{-39}$ for HBL--LBL; 
$5 \times 10^{-40}$ for HBL--FSRQ; 
$0.7$  for LBL--FSRQ.
\label{fig:alpha_rx} }
\end{figure}

\subsection{Histograms}
\label{sec:histograms}

\begin{table}
\begin{center}
\begin{tabular}{|l|ccc|}
\hline
&HBL&LBL&FSRQ\\
\hline
$\alpha_{x}$[2--10 keV]    &   1.34$\pm$0.05  &  0.84$\pm$0.07 &  0.65$\pm$0.04\\
$\alpha_{x}$[0.1--2.4 keV] &   1.28$\pm$0.04  &  1.39$\pm$0.08 &  0.76$\pm$0.06\\
$z$                        &   0.25$\pm$0.02  &  0.46$\pm$0.05 &  1.27$\pm$0.12\\
Log $\nu_{r}L_{\nu_{r}}$   &  41.51$\pm$0.07  & 43.65$\pm$0.16 & 44.93$\pm$0.13\\
Log $\nu_{o}L_{\nu_{o}}$   &  44.66$\pm$0.06  & 45.49$\pm$0.12 & 46.35$\pm$0.11\\
Log $\nu_{x}L_{\nu_{x}}$   &  44.62$\pm$0.08  & 44.52$\pm$0.15 & 45.89$\pm$0.11\\
$\alpha_{ro}$              &   0.36$\pm$0.01  &  0.56$\pm$0.02 &  0.66$\pm$0.01\\
$\alpha_{ox}$              &   1.03$\pm$0.02  &  1.38$\pm$0.02 &  1.21$\pm$0.02\\
$\alpha_{rx}$              &   0.59$\pm$0.01  &  0.84$\pm$0.01 &  0.85$\pm$0.01\\
\hline
\end{tabular}
\caption{Average values of X--ray spectral indices, redshifts,
$\nu L_\nu$ luminosities in different bands and broad band spectral indices.
The listed errors are weighted errors.
}
\end{center}
\label{tab:three}
\end{table}

The distribution of spectral indices, redshifts and luminosities are
shown by the histograms in Fig. 1 -- Fig. 9.
In the figure captions we give the probability, according to the 
Kolmogorov--Smirnov (KS) test, that two distributions are drawn from 
the same parent population, comparing HBLs and LBLs, HBLs and FSRQs,
LBLs and FSRQs.
%

The mean values of the plotted quantities are listed in Table 3.

\subsubsection{Spectral indices}

The distributions of the energy spectral indices (Fig.~\ref{fig:alpha_hard}
and Fig.~\ref{fig:alpha_soft}) show that for FSRQ we have 
an average value less than unity in both energy ranges.
This suggests that for this subclass of blazars we are observing only the 
inverse Compton component in the entire X--ray band, from 0.1 to 10 keV.
For HBL, instead, the average energy spectral index is greater than 
unity, indicating
that we are observing the synchrotron component after its peak. 
On average, LBL show a flattening going from the soft to the hard 
X--ray bands.

These results suggest that both the soft and the hard X--ray bands are 
dominated by the inverse Compton process in FSRQs and by the synchrotron
process in HBL, while in LBL we have the synchrotron flux dominating
in the soft band and the flatter Compton component emerging at higher X--ray
energies.

\subsubsection{Redshift}

While the redshifts of FSRQs are quite uniformly distributed up to a value
of $\simeq$3, BL Lacs have redshifts lower than 1 (and HBL have lower
redshifts than LBL, see Fig.~\ref{fig:z}). 
There is no significant difference between the redshift distributions of 
sources observed in the hard and in the soft X--ray bands.
Of the sources in our sample, about 25\% have no measured redshift
(38 HBL and 17 LBL).
This incompletness, even if not severe, could bias the shown redshift
distribution of HBLs towards the lower part (since larger
redshifts are more difficult to measure).

\subsubsection{Luminosities}

From the radio, optical and 1 keV monochromatic fluxes we have calculated
the ``$\nu L_\nu$'' luminosities in the corresponding bands.  
The distributions of radio and optical luminosities (Fig.~\ref{fig:lr} 
and Fig.~\ref{fig:lo}) show a continue variation in the three subclasses of
blazar: HBLs are the least powerful sources, and FSRQs are the most
luminous objects. 
this is more pronounced in the radio than in the optical band.
The X--ray luminosities of HBLs and LBLs are very similar
(Fig.~\ref{fig:lx}), while FSRQs are more luminous by a factor of 10.

\subsubsection{Broad band indices}

Also the broad band spectral index $\alpha_{ro}$ changes smoothly
between the subclasses of blazar (Fig.~\ref{fig:alpha_ro}). 
On average, it  becomes steeper going from HBL to FSRQs. The optical to
X--ray broad band index distribution (Fig.~\ref{fig:alpha_ox}) is broader
for HBL, with an average value smaller than for LBL and FSRQs.
The spectral index $\alpha_{rx}$ (Fig.~\ref{fig:alpha_rx}) is on average
the same for FSRQs and LBLs, and obviously flatter (by definition) for HBLs.


\section{The average SED}
\label{sec:average_sed}

\begin{figure*}
\vspace{-1cm}
\psfig{figure=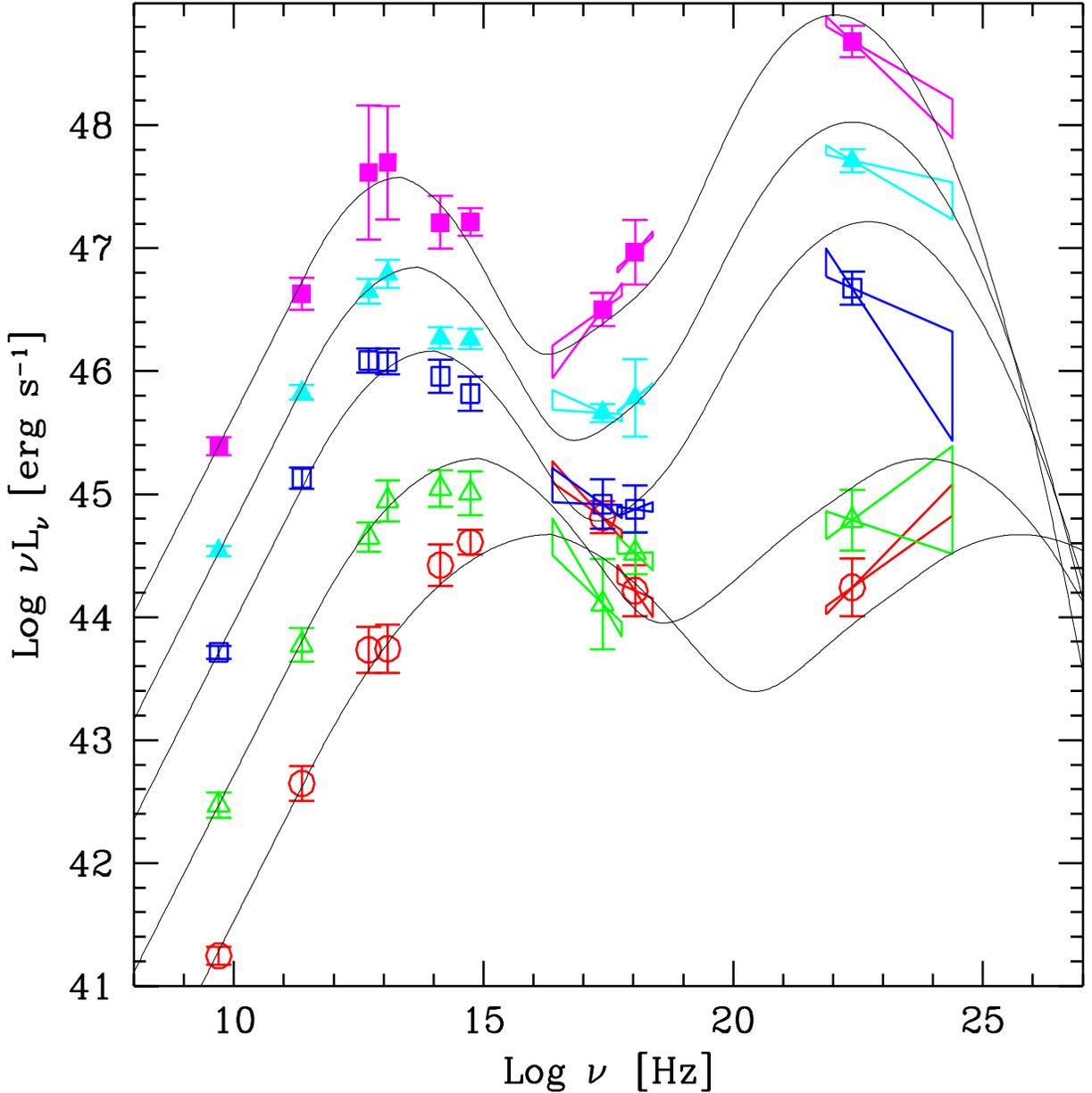,width=19.3cm,height=20cm}
\vspace{-1.5cm}
\caption{The average SED of the blazars studied by Fossati et al. (1998),
including the average values of the hard X--ray spectra.
The thin solid lines are the spectra constructed following the 
parameterization proposed in this paper.
\label{fig:avrg_seds} }
\end{figure*}

In Fig.~\ref{fig:avrg_seds} we show the sequence of average SEDs as
published by F98, but including the [2--10 keV] averages spectral indices
and fluxes.  The latter have been constructed considering only
the same samples considered by F98.

It can be seen that in general the 2--10 keV fluxes and spectral indices
connect smoothly on the softer {\it ROSAT} data even if they are,
on average, flatter than the latter.
This is due to the emergence, in the hard X--ray band, of the inverse
Compton component which is progressively more dominant as the luminosity
increases.

For the average SED corresponding to the second lowest luminosity bin,
there is a mismatch between the soft and hard X--ray data.
By comparing the data of each source in common, we found that 
all 5 sources were bright when observed by ASCA or {\it Beppo}SAX
than at the time of the {\it ROSAT} observations.
We therefore believe that the mismatch is due to the variable nature of the
objects and the small number of sources in this luminosity bin.

The average spectral indices of the objects in common with F98 are listed
in Table~\ref{tab:averages}, which also lists the average luminosities
at 4.47 keV (the logarithmic mid point between 2 and 10 keV).

\begin{table}
\begin{center}
\begin{tabular}{|cccc|}
\hline
$<\log\nu_r L_{\nu_r}>$ &$<\log\nu_xL_{\nu_x}>$ &N$_{\rm sources}$ &$\alpha_x$ \\     
                  & @4.47 keV  & & 2--10 keV   \\
\hline
$<$ 42       & 44.2      &   12   &1.39$\pm$0.21  \\
$42$--$43$   & 44.5      &    5   &1.19$\pm$0.21   \\
$43$--$44$   & 44.9      &    6   &0.95$\pm$0.11   \\
$44$--$45$   & 45.8      &    6   &0.68$\pm$0.02   \\
$>$45        & 47.0      &   11   &0.58$\pm$0.06   \\
\hline
\end{tabular}
\caption{Average values of the X--ray luminosity at 4.47 keV ($\nu L_\nu$
values) and average 2--10 keV spectral indices, for the sources in common with 
Fossati et al. 1998, for each radio luminosity bin.}
\end{center}
\label{tab:averages}
\end{table}
%
	
The continuous lines in Fig.~\ref{fig:avrg_seds} correspond to a simple
parametric model derived by the one introduced by Fossati et al. (1998).
We introduce minor modifications, adopted both to better represent our data
at small luminosities and to follow a more physical scenario, in which the
low power HBLs can be described by a pure synchrotron--self Compton model
(see e.g. Ghisellini et al., 1998).  
We remind the reader here of the key assumptions of the F98 parametric model:
\begin{itemize}

\item[$\bullet$] The observed radio luminosity $L_R=(\nu L_\nu)|_{5GHz}$ 
is assumed to be linearly proportional to the bolometric luminosity, 
and related to the location of the synchrotron peak through:
\begin{equation}
\nu_{s} \, \propto L_R^{-\eta}
\end{equation}
where $\eta=1.8$ for $L_R<3\times 10^{42}$ erg s$^{-1}$ and
$\eta= 0.6$ for $L_R>3\times 10^{42}$ erg s$^{-1}$.

\item[$\bullet$] The ratio between the Compton and the synchrotron peak
frequencies is constant: $\nu_c/\nu_s = 5\times 10^8$ for all luminosities.

\item[$\bullet$] The ratio between the power of the inverse Compton and the
radio powers is constant: $L_c/L_R = 3\times 10^3$ for all luminosities.

\item[$\bullet$] The ratio between the radio and X--ray (at 1 keV) Compton
luminosity is fixed.

\end{itemize}
The SED is then constructed assuming for the synchrotron component
a flat ($\propto \nu^{-0.2}$) radio spectrum connecting to a parabola 
(in log--log space) peaking at $\nu_s$.
The junctions between the power law and the parabola is continuous.
For the inverse Compton spectrum it is assumed that an initial power law
of index $\alpha = 0.6$ ends in another parabola peaking at $\nu_c$.


\noindent
We modified the Fossati et al. (1998) description in the following way:

\begin{itemize}
\item[$\bullet$] We changed the values of $\eta$, assuming $\eta=1.2$ and
0.4 for $L_R$ smaller and greater than $10^{43}$ erg s$^{-1}$.

\item[$\bullet$] The ratio $\nu_c/\nu_s$ is assumed to be constant with the
same value as before for $L_R>10^{43}$ erg s$^{-1}$, but for smaller radio
luminosity we set:
\begin{equation}
{\nu_c \over \nu_s} \,  = \, 5\times 10^8 L_{R,43}^{-0.4}
\end{equation}

\item[$\bullet$] Below $L_R<10^{43}$ erg s$^{-1}$ we assume that the
synchrotron and Compton peaks have the same luminosities.
For greater $L_R$ we assumed, as before, $L_c/L_R=3\times 10^3$.
\end{itemize}

\noindent
The spectra predicted by this new parameterization are shown in
Fig.~\ref{fig:avrg_seds} as thin solid lines.
As anticipated, the assumptions described above have a physical motivation.
In fact, for low luminosity sources, we have evidences that the seed
photons producing the Compton spectrum are the locally produced synchrotron
ones, with no or negligible contributions from seed photons produced
externally to the jet (e.g. from the Broad Line Region).
In this case:
\begin{itemize}

\item[i)]  The ratio $\nu_c/\nu_s$ increases with $\nu_s$ as long as 
the scattering process is in the Thomson regime, and decreases
with $\nu_s$ in the Klein Nishina regime.

\item[ii)] On average, the BL Lacertae objects detected by EGRET with
$L_R<10^{43}$ erg s$^{-1}$ have roughly the same synchrotron and Compton
components.

\item[iii)] The radio luminosity $L_R=10^{43}$ erg s$^{-1}$
may corresponds to the power
for which emission lines and/or external seed photons becomes important for
the formation of the inverse Compton spectrum 
(see e.g. Ghisellini et al., 1998).
\end{itemize}

\begin{figure}
\psfig{figure=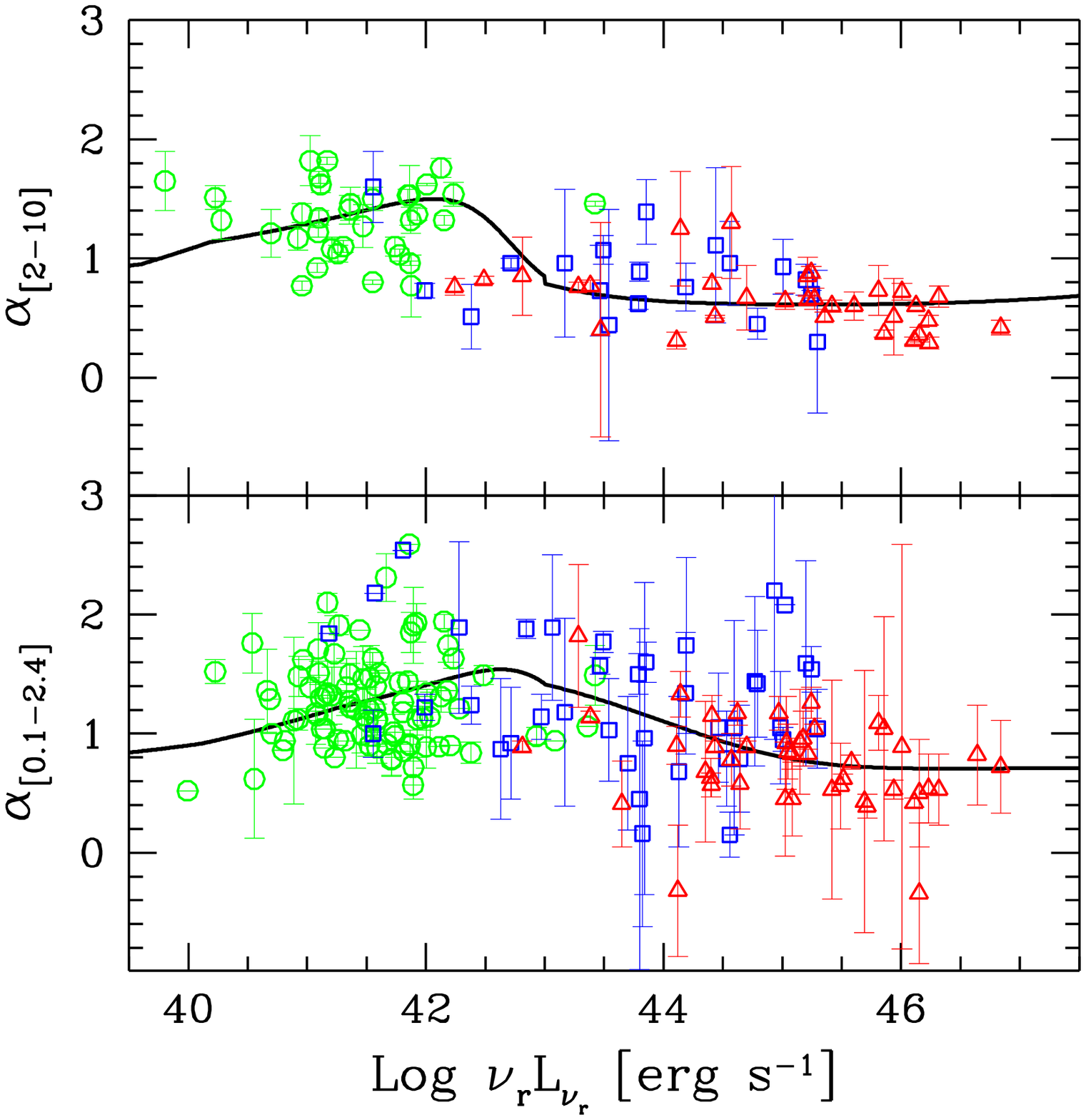,width=9.5cm}
\vspace{-0.5cm}
\caption{Soft and hard X--ray spectral index vs. the radio luminosity.
Circles: HBL, Squares: LBL, Triangles: FSRQs.
\label{fig:ax_lr} }
\end{figure}

\begin{figure}
\psfig{figure=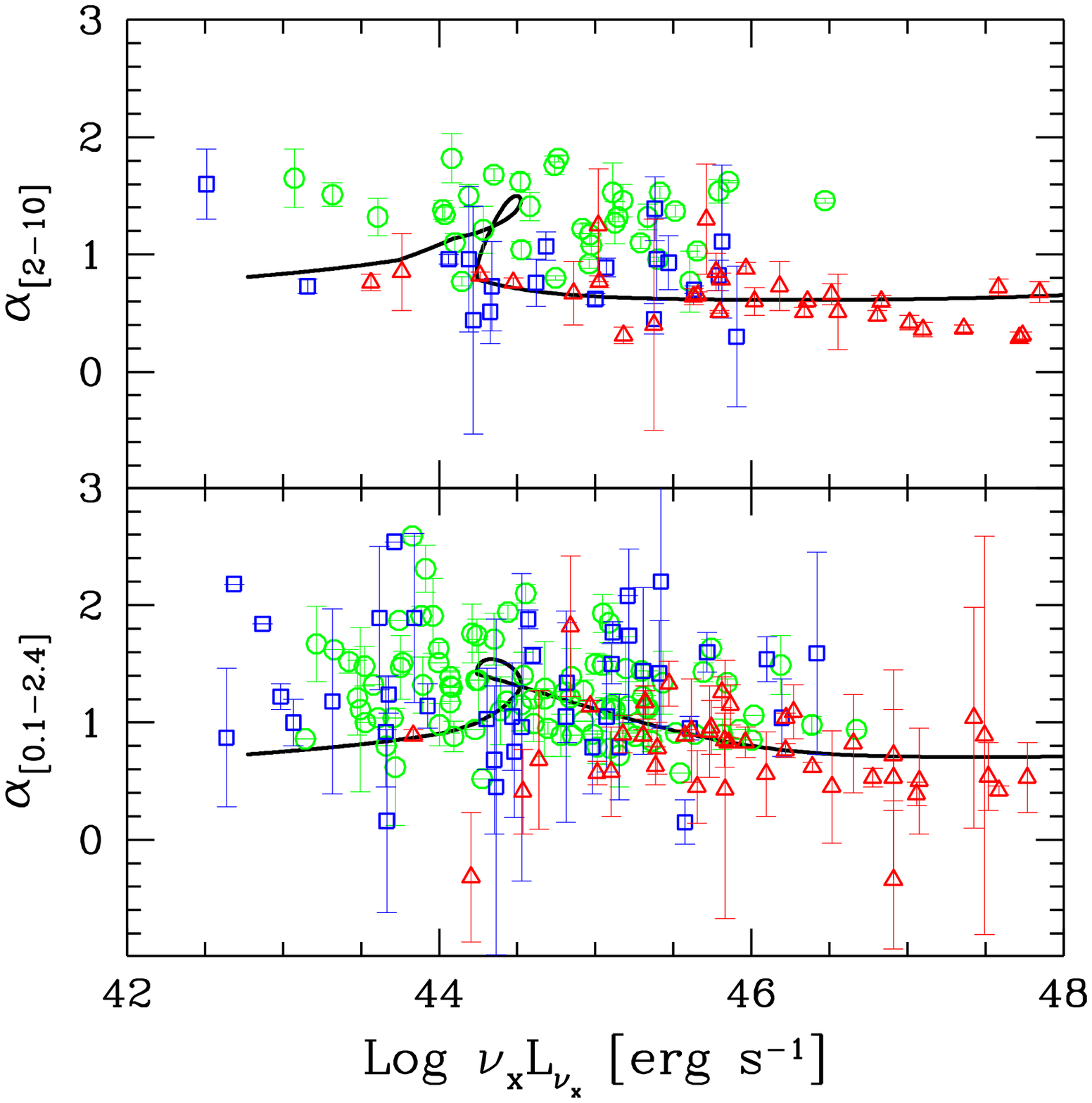,width=9.5cm}
\vspace{-0.5cm}
\caption{Soft and hard X--ray spectral index vs. the X--ray luminosity.
Circles: HBL, Squares: LBL, Triangles: FSRQs.
\label{fig:ax_lx} }
\end{figure}

\begin{figure}
\psfig{figure=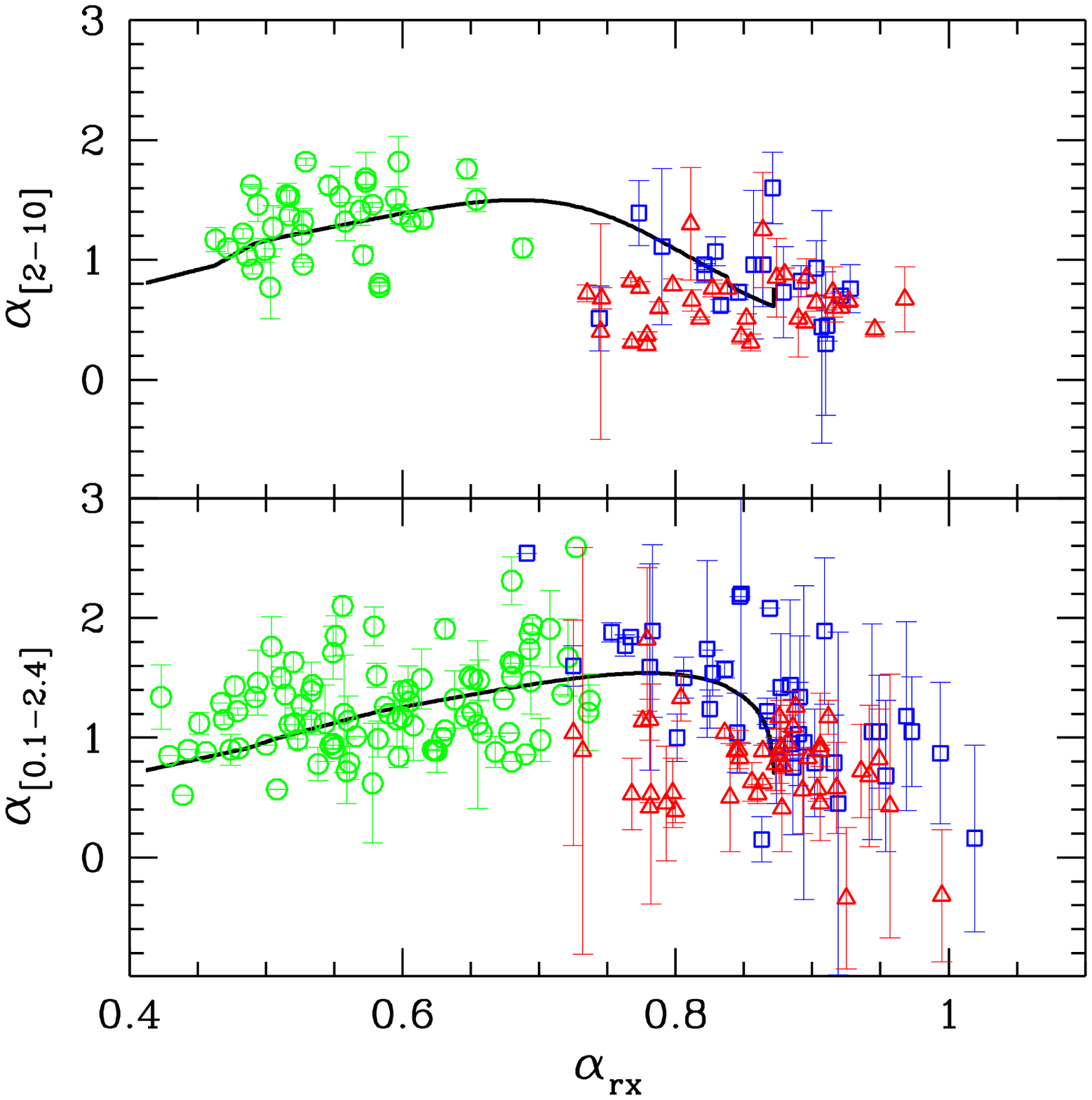,width=9.5cm,height=11.5cm}
\vskip -1 true cm
\caption{Soft and hard X--ray spectral index vs. the broad band radio to
X--ray index.  Circles: HBL, Squares: LBL, Triangles: FSRQs.
\label{fig:ax_arx} }
\end{figure}

\begin{figure}
\psfig{figure=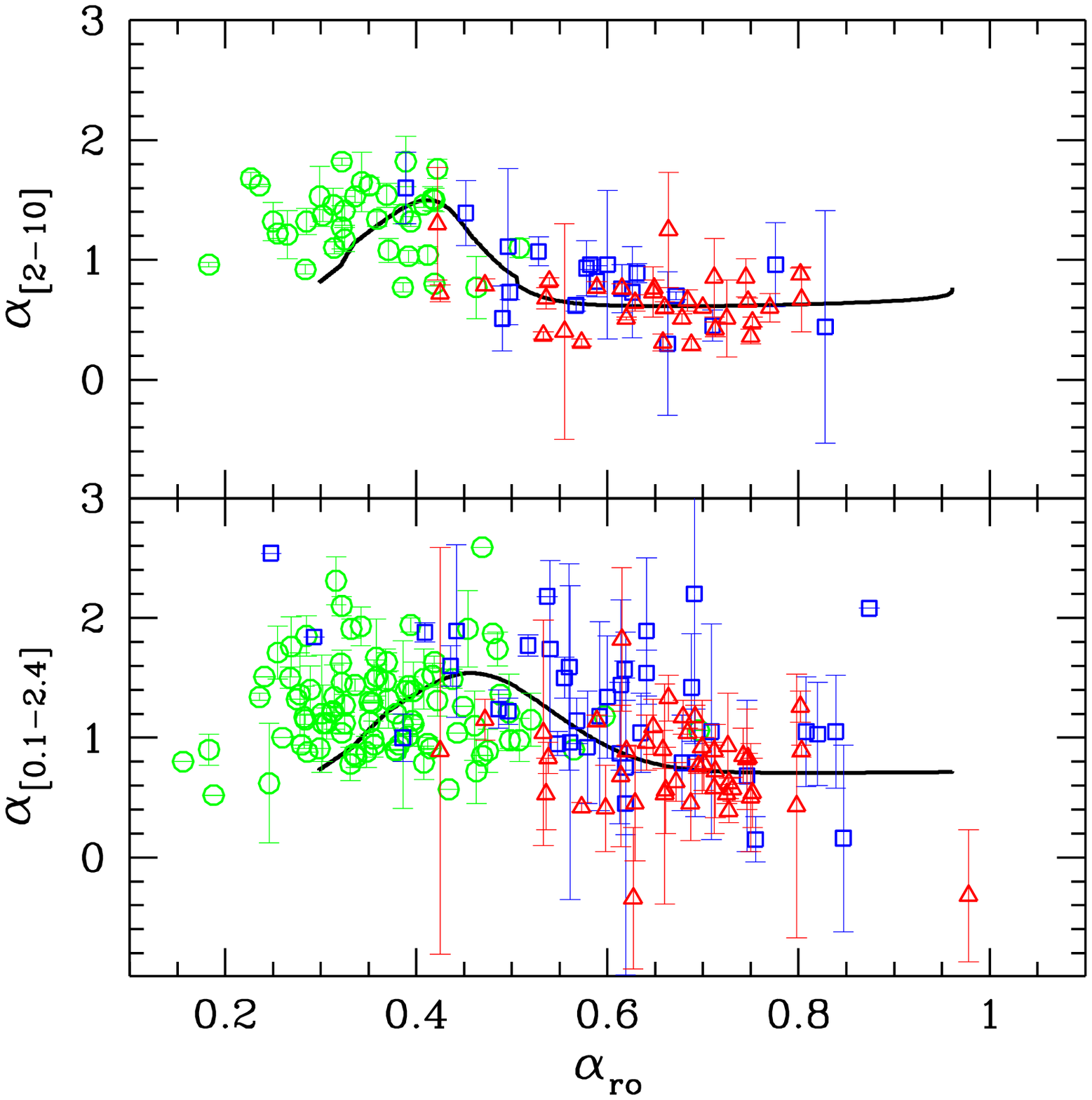,width=9.5cm,height=11.5cm}
\vskip -1 true cm
\caption{Soft and hard X--ray spectral index vs. the broad band radio to
optical index.  Circles: HBL, Squares: LBL, Triangles: FSRQs.
\label{fig:ax_aro} }
\end{figure}

\begin{figure}
\psfig{figure=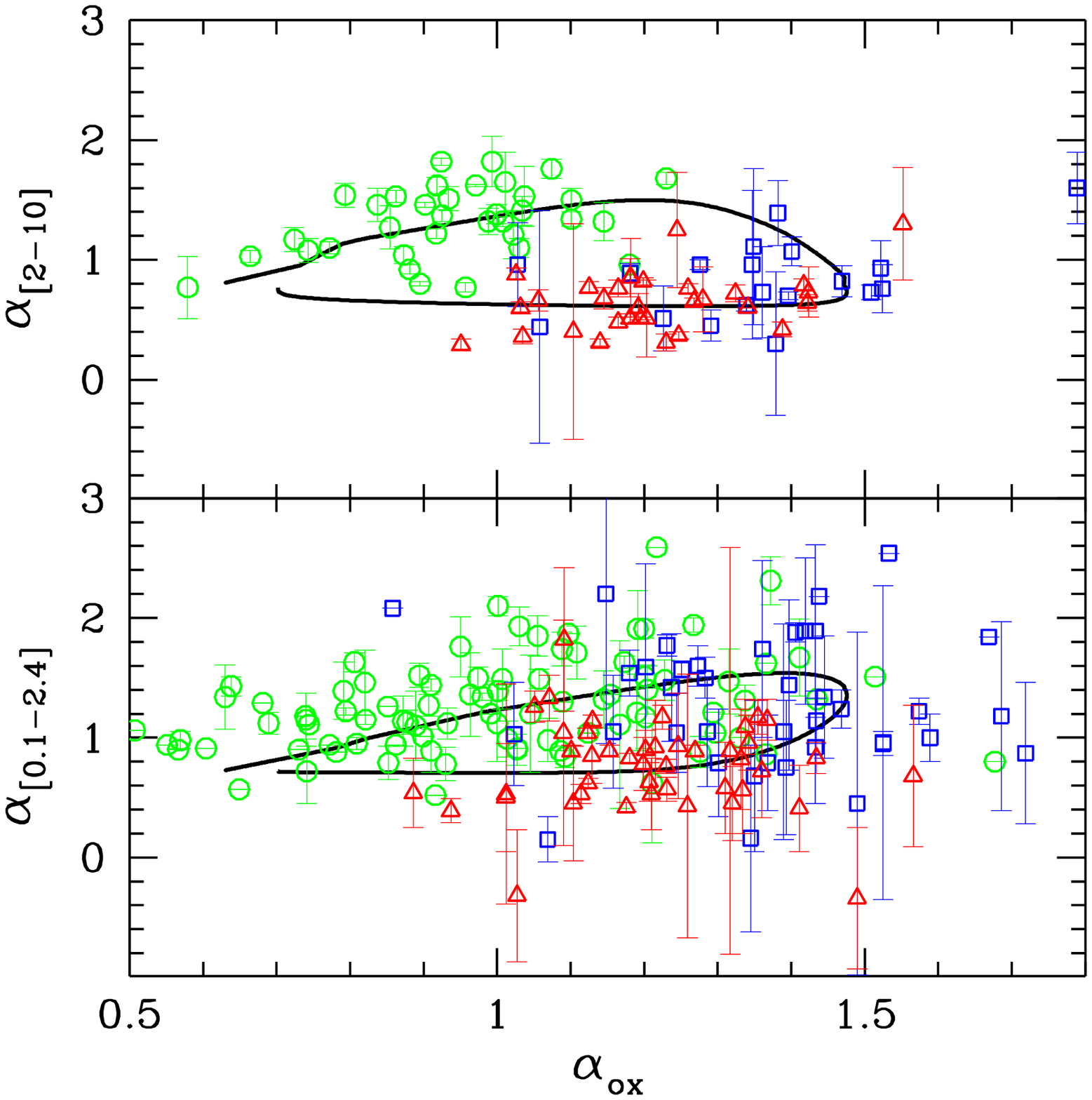,width=9.5cm,height=11.5cm}
\vskip -1 true cm
\caption{Soft and hard X--ray spectral index vs. the broad band optical to
X-ray index.  Circles: HBL, Squares: LBL, Triangles: FSRQs.
\label{fig:ax_aox} }
\end{figure}

\begin{figure}
\psfig{figure=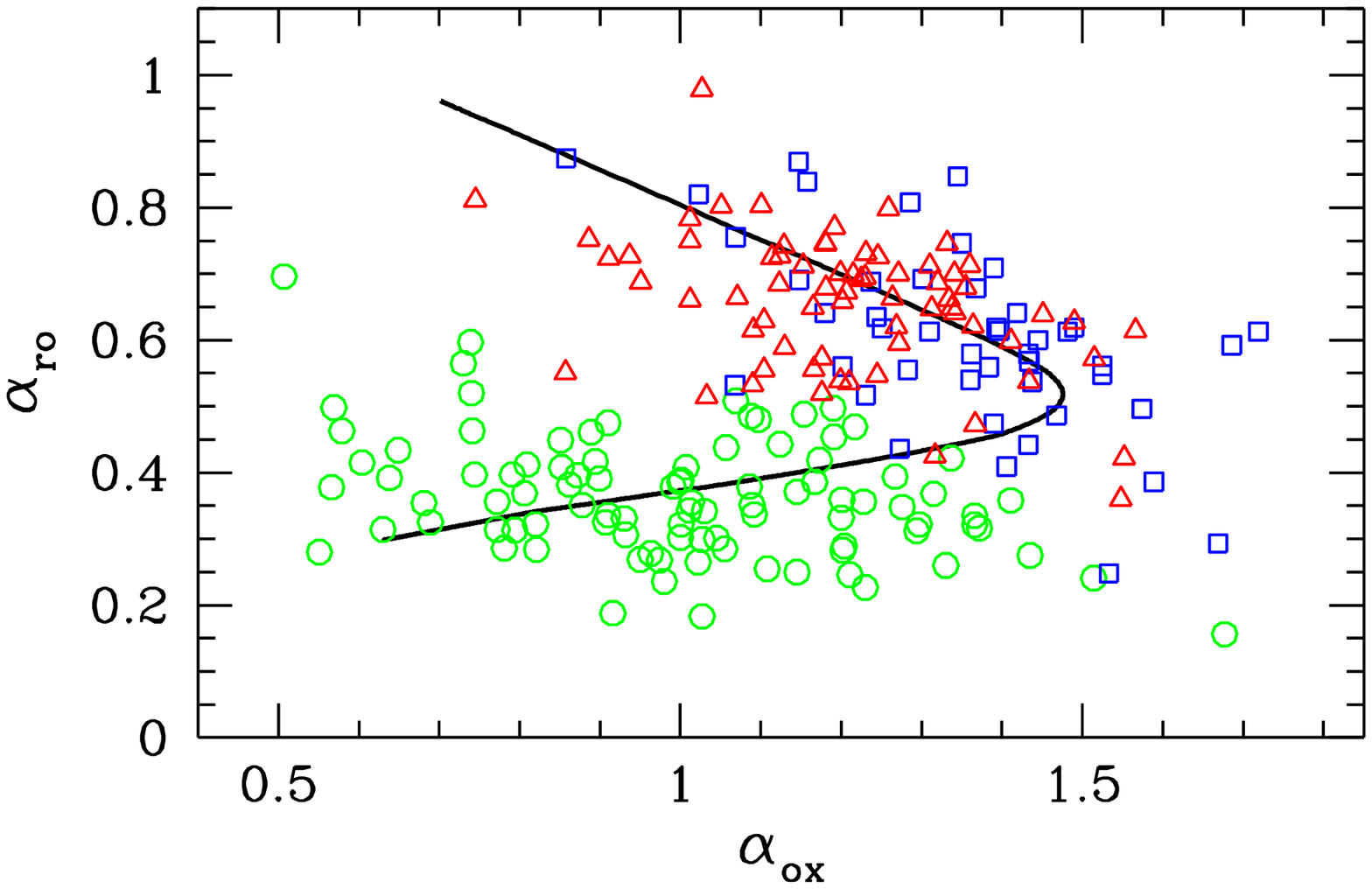,width=9.5cm,height=11.5cm}
\vskip -4 true cm
\caption{Radio to optical vs. optical to X--ray broad band indices.
Circles: HBL, Squares: LBL, Triangles: FSRQs.
\label{fig:aox_aro} }
\end{figure}

\begin{figure}
\psfig{figure=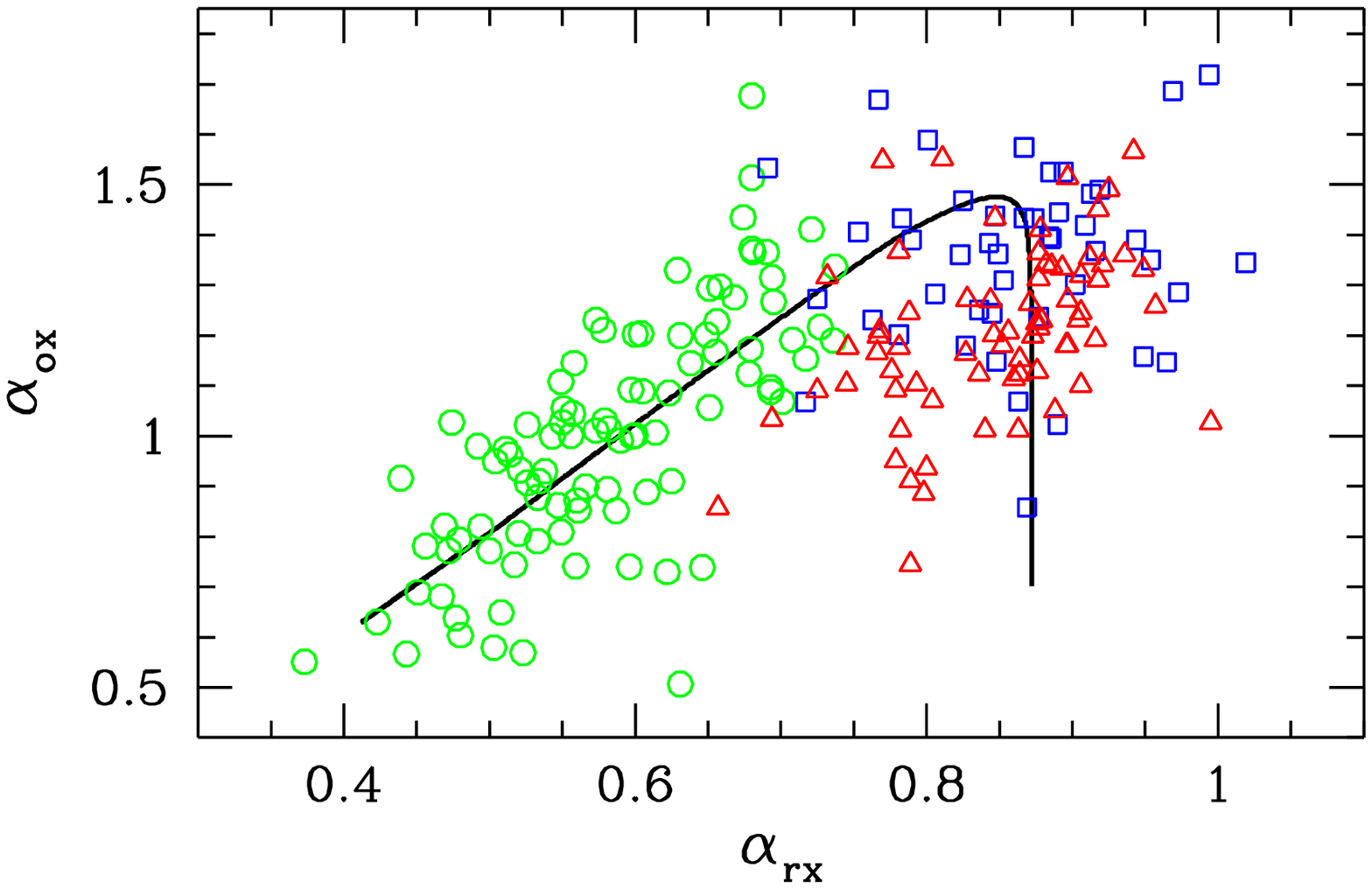,width=9.5cm,height=11.5cm}
\vskip -4 true cm
\caption{Optical to X--ray vs. radio to X--ray broad band indices.
Circles: HBL, Squares: LBL, Triangles: FSRQs.
\label{fig:arx_aox} }
\end{figure}

\begin{figure}
\psfig{figure=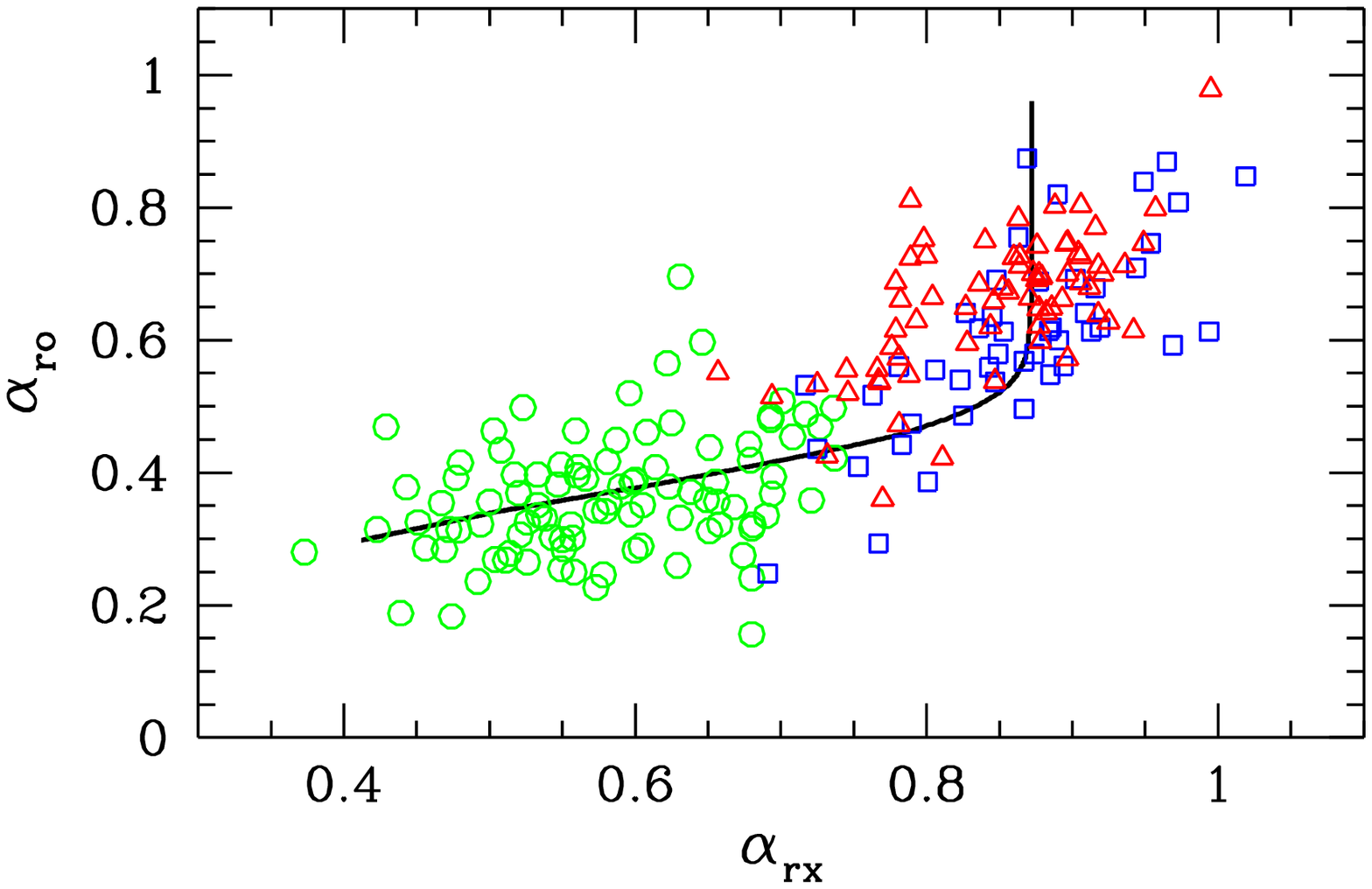,width=9.5cm,height=11.5cm}
\vskip -4 true cm
\caption{Radio to optical vs. radio to X--ray broad band indices.
Circles: HBL, Squares: LBL, Triangles: FSRQs.
\label{fig:arx_aro} }
\end{figure}

\section{Correlations}
\label{sec:correlations}

In order to see if the proposed parameterization of the average SED of 
blazars (constructed on the basis of a subsample of the sources 
in our catalogue) can account for the general properties of all
the blazars in our catalogue, we have investigated the correlations
between the X--ray spectral index (both hard and soft) with the 
radio and X--ray luminosities and with the broad band spectral indices.
We have then considered the correlations between broad band indices.
The results are shown in Fig.~\ref{fig:ax_lx}--\ref{fig:arx_aro} (solid
line), where we have superposed the relations predicted by the new
parameterization.

In Fig.~\ref{fig:ax_lr} and Fig.~\ref{fig:ax_lx} we show the hard and soft
X--ray indices as a function of the radio and X--ray luminosity for all
sources, and compare these data with the model expectations.
Note that the latter have been constructed to describe the {\it average} 
properties of blazars, whatever the scatter around it.
Bearing this in mind, we can consider the description of the model
quite satisfactorily.

In Fig.~\ref{fig:ax_arx} we show the hard and soft X--ray indices as a
function of the broad band index $\alpha_{rx}$.
The model well describes the small $\alpha_{rx}$ part (corresponding to HBLs), 
but fails to describe the X--ray flattest sources with the steeper
value of $\alpha_{rx}$.
This is due to two reasons:
i) the slope of the power law of the
Compton component is assumed to be $\alpha=0.7$, so that flatter
indices are not possible;
ii) the ratio between the radio and the 1 keV Compton luminosity
is fixed for all sources. 
This results in a saturated value of $\alpha_{rx}\sim 0.85$,
occurring when the Compton component dominates at 1 keV
(i.e. for powerful sources).

In Fig.~\ref{fig:ax_aro} and Fig.~\ref{fig:ax_aox} we show $\alpha_{X}$  as
a function of $\alpha_{ro}$ and $\alpha_{ox}$: the average properties are
well described by the model.

In Fig.~\ref{fig:aox_aro}, we show the ``classic''
$\alpha_{ro}$--$\alpha_{ox}$ diagram for the two subsamples of sources.
Note that the model well describes all the data but those sources
with the largest values of $\alpha_{ox}$, corresponding to ``transition''
sources between the LBLs and FSRQ.

Fig.~\ref{fig:arx_aox}, Fig.~\ref{fig:arx_aro}: note that HBL are well
separated from the other classes of blazars. 
Note also that the model, by construction, 
has an asymptotic limit for $\alpha_{rx}$, whose value therefore 
saturates at $\alpha_{rx} \simeq 0.85$.

\section{Discussion}
\label{sec:discussion}

The large database of X--ray spectra of blazars we have collected
has allowed to test the blazar sequence scheme aiming to
unify the different behaviors of blazars on the basis of a single
parameter, i.e. the bolometric observed luminosity.
We have found that the proposed parameterization can account for the average
properties of the blazars in our sample, even if the scatter around the
predicted average quantities is sometimes large.

We confirm, on a statistical basis, that more powerful blazars
emit the X--rays by the inverse Compton process, while in less powerful
objects the dominant mechanism is synchrotron, and the transition
is smooth, with LBL possibly showing both contributions.

The Fossati et al. (1998) parameterization scheme to reproduce
the blazar SED is able to fit also our new hard X--ray data and the 
correlations between broad band indices of the sources in our sample.
However, we propose to slightly change this parameterization
especially for the low power objects (i.e. the HBLs), in line with
the idea that HBLs are characterized
by a pure synchrotron self--Compton spectrum, without extra contributions
produced by non locally produced seed photons.

In this paper, and in Fossati et al. (1998), the power law relation between 
the synchrotron peak frequency and the radio luminosity changes
slope at some critical radio power, of the order of 3--10$\times 10^{42}$
erg s$^{-1}$.
This agrees with the absence of broad emission lines in these objects.
It has still to be proven, however, if the non visibility of emission 
lines in low power objects is due to a genuine lack of emitting clouds, 
or is due to a weak ionizing continuum.
If the latter hypothesis is true, then we expect that the broad line 
region indeed exists, but at smaller radii than in more powerful objects.
In this case, the lack of the external Compton component is not due
to the lack of external photons, but possibly to the fact that
the dissipation region in these sources is beyond the broad line region:
in this case the corresponding energy density of line photons
is seen, in the comoving frame of the blob, depressed by the square of
the bulk Lorentz factor.

\acknowledgements{We thank the anonimous referee for useful
suggestions and Luigi Costamante for discussions.
This research made use of the NASA/IPAC Extragalactic Database (NED)
which is operated by the Jet Propulsion Laboratory, Caltech, under
contract with the National Aeronautics and Space Administration.
We acknowledge finantial support from the MURST.
}


\begin{table*}
\begin{tabular}{lccccccccccccc}
\hline
IAU Name& z & $F_{\rm R}$ & Ref. & $F_{\rm O}$ & Ref. & $F_{\rm X}$ & Ref. & $\Gamma$ & err & Ref. & Type & Ref. & Sat \\
        &   &   (Jy)      &      &   (mJy)     &      &  ($\mu$Jy)  &      &          &     &      &      &      &     \\
\hline \hline 
0014$+$813&   3.366&    0.551&  Ca97 &    0.910&  Ca97 &    0.350&  Ca97 &    1.890&  1.700&  Ca97 &     3&  Ca97 & RO \\
          &        &         &       &         &       &    0.430&  Ca97 &    1.720&  0.070&  Ca97 &      &       & AS \\
0016$+$731&   1.781&    1.750&  NED  &    0.060&  NED  &    0.050&  Sa97b&    1.430&  1.100&  Sa97b&     3&  Gh93 & RO \\
0038$-$020&   1.178&    0.581&  NED  &    0.030&  NED  &    0.100&  Wo90 &    1.400&  1.100&  Wo90 &     3&  Wo90 & EI \\
0045$+$214&     ...&    0.043&  NED  &    0.310&  NED  &    0.450&  Br97 &    2.160&  0.250&  Br97 &     1&  Br97 & RO \\
0048$-$097&   0.200&    2.743&  Fo98 &    1.750&  Fo98 &    0.420&  Pa98 &    1.730&  0.380&  Pa98 &     2&  Gh93 & SA \\
\end{tabular}
\label{tab:listone}
\end{table*}
\begin{table*}
\begin{tabular}{lccccccccccccc}
\hline
IAU Name& z & $F_{\rm R}$ & Ref. & $F_{\rm O}$ & Ref. & $F_{\rm X}$ & Ref. & $\Gamma$ & err & Ref. & Type & Ref. & Sat \\
        &   &   (Jy)      &      &   (mJy)     &      &  ($\mu$Jy)  &      &          &     &      &      &      &     \\
\hline \hline 
0350$-$3712&  0.165&    0.017&  VV93 &    0.590&  NED  &    0.350&  La96 &    2.170&  0.260&  La96 &     1&  La96 & RO \\
0403$-$132&   0.571&    2.889&  Fo98 &    0.540&  Fo98 &    0.400&  Fo98 &    1.780&  0.220&  Fo98 &     3&  IT90 & RO \\
          &        &         &       &         &       &    0.490&  Wo90 &    4.300&  4.300&  Wo90 &      &       & EI \\
0405$-$123&   0.574&    1.960&  Fo98 &    5.170&  Fo98 &    0.090&  TW   &    1.760&  0.090&  TW   &     3&  Sa97b& AS \\
          &        &         &       &         &       &    1.160&  Fo98 &    2.150&  0.170&  Fo98 &      &       & RO \\
0406$+$121&   1.020&    0.746&  Na96 &    0.120&  NED  &    0.050&  La96 &    1.790&  0.450&  La96 &     2&  La96 & RO \\
\end{tabular}
\end{table*}
\begin{table*}
\begin{tabular}{lccccccccccccc}
\hline
IAU Name& z & $F_{\rm R}$ & Ref. & $F_{\rm O}$ & Ref. & $F_{\rm X}$ & Ref. & $\Gamma$ & err & Ref. & Type & Ref. & Sat \\
        &   &   (Jy)      &      &   (mJy)     &      &  ($\mu$Jy)  &      &          &     &      &      &      &     \\
\hline \hline 
0737$+$746&   0.315&    0.024&  Fo98 &    0.650&  Fo98 &    0.880&  Wo98 &    2.530&  0.250&  Wo98 &     1&  Wo98 & SA \\
          &        &         &       &         &       &    1.110&  Fo98 &    1.910&  ---  &  Fo98 &      &       & RO \\
0754$+$100&   0.660&    1.480&  Pa95 &    3.640&  Pa95 &    0.720&  Sa94 &    2.110&  0.650&  Sa94 &     2&  Sa94 & EX \\
0804$+$499&   1.433&    2.050&  Co97 &    0.390&  Co97 &    0.170&  Co97 &    1.560&  0.360&  Co97 &     3&  Gh93 & RO \\
0806$+$3504&   0.082&    0.180&  NED  &    0.660&  NED &    0.740&  Ba98 &    2.870&  ---  &  Ba98 &     1&  Ba98 & RO \\
0806$+$524&   0.138&    0.177&  NED  &    3.150&  NED  &    4.910&  Br94 &    2.930&  0.160&  Br94 &     1&  Br94 & RO \\
\end{tabular}
\end{table*}
\begin{table*}
\begin{tabular}{lccccccccccccc}
\hline
IAU Name& z & $F_{\rm R}$ & Ref. & $F_{\rm O}$ & Ref. & $F_{\rm X}$ & Ref. & $\Gamma$ & err & Ref. & Type & Ref. & Sat \\
        &   &   (Jy)      &      &   (mJy)     &      &  ($\mu$Jy)  &      &          &     &      &      &      &     \\
\hline \hline 
1055$+$567&   0.144&    0.260&  NED  &    1.810&  NED  &    0.470&  Br97 &    2.310&  0.130&  Br97 &     1&  Br97 & RO \\
          &        &         &       &         &       &    2.190&  Sb99 &    2.760&  0.080&  Sb99 &      &       & AS \\
          &        &         &       &         &       &    1.080&  Sb99 &    2.780&  0.250&  Sb99 &      &       & SA \\
1101$+$384&   0.031&    0.722&  Fo98 &   16.970&  Fo98 &   58.370&  Fo98b&    2.820&  0.030&  Fo98b&     1&  Gh93 & SA \\
          &        &         &       &         &       &   23.850&  Ku98 &    2.970&  0.020&  Ku98 &      &       & AS \\
          &        &         &       &         &       &   36.100&  Fo98 &    3.100&  0.080&  Fo98 &      &       & RO \\
\end{tabular}
\end{table*}
\begin{table*}
\begin{tabular}{lccccccccccccc}
\hline
IAU Name& z & $F_{\rm R}$ & Ref. & $F_{\rm O}$ & Ref. & $F_{\rm X}$ & Ref. & $\Gamma$ & err & Ref. & Type & Ref. & Sat \\
        &   &   (Jy)      &      &   (mJy)     &      &  ($\mu$Jy)  &      &          &     &      &      &      &     \\
\hline \hline 
1255$+$244&   0.141&    0.077&  Pa95 &    2.510&  Pa95 &    1.450&  Gr96 &    2.400&  0.200&  Gr96 &     1&  Gr96 & RO \\
1303$+$5056&  0.688&    0.001&  Ba98 &    0.030&  NED  &    0.830&  Ba98 &    1.940&  ---  &  Ba98 &     1&  Ba98 & RO \\
1308$+$326&   0.997&    1.832&  Fo98 &    1.580&  Fo98 &    0.110&  TW   &    1.990&  0.420&  TW   &     2&  Gh93 & AS \\
          &        &         &       &         &       &    0.090&  TW   &    1.740&  0.250&  TW   &      &       & AS \\
          &        &         &       &         &       &    0.150&  Co95 &    1.950&  0.100&  Co95 &      &       & RO \\
1312$-$423&   0.105&    0.019&  Pa95 &    0.830&  Pa95 &    1.520&  Wo98 &    2.210&  0.200&  Wo98 &     1&  Wo98 & SA \\
\end{tabular}
\end{table*}
\begin{table*}
\begin{tabular}{lccccccccccccc}
\hline
IAU Name& z & $F_{\rm R}$ & Ref. & $F_{\rm O}$ & Ref. & $F_{\rm X}$ & Ref. & $\Gamma$ & err & Ref. & Type & Ref. & Sat \\
        &   &   (Jy)      &      &   (mJy)     &      &  ($\mu$Jy)  &      &          &     &      &      &      &     \\
\hline \hline 
1542$+$614&     ...&    0.131&  NED  &    0.960&  NED  &    0.120&  LM99 &    2.500&  ---  &  LM99 &     2&  LM99 & RO \\
1546$+$027&   0.413&    1.450&  Kh81 &    0.230&  NED  &    0.840&  Wo90 &    2.180&  2.250&  Wo90 &     3&  Wo90 & EI \\
1548$+$114&   0.436&    0.410&  VV93 &    0.500&  Gh93 &    0.320&  Wo90 &    1.350&  1.100&  Wo90 &     3&  Gh93 & EI \\
1552$+$2020&  0.222&    0.038&  Pa95 &    0.300&  Pa95 &    1.570&  Pe96 &    1.790&  0.140&  Pe96 &     1&  Pe96 & RO \\
1553$+$113&   0.360&    0.636&  Pa95 &    4.800&  Pa95 &   13.420&  BSMc &    2.850&  0.040&  BSMc &     1&  NED  & SA \\
          &        &         &       &         &       &   14.860&  TW   &    2.470&  0.180&  TW   &      &       & AS \\
\end{tabular}
\end{table*}
\begin{table*}
\begin{tabular}{lccccccccccccc}
\hline
IAU Name& z & $F_{\rm R}$ & Ref. & $F_{\rm O}$ & Ref. & $F_{\rm X}$ & Ref. & $\Gamma$ & err & Ref. & Type & Ref. & Sat \\
        &   &   (Jy)      &      &   (mJy)     &      &  ($\mu$Jy)  &      &          &     &      &      &      &     \\
\hline \hline 
1848$+$427&     ...&    0.007&  Br97 &    0.200&  NED  &    1.810&  Br94 &    2.170&  0.220&  Br94 &     1&  Br94 & RO \\
1921$-$293&   0.352&   12.696&  NED  &    0.840&  NED  &    0.770&  Wo90 &    1.570&  1.300&  Wo90 &     3&  Gh93 & EI \\
          &        &         &       &         &       &    0.380&  Go95 &    1.670&  0.270&  Go95 &      &       & EX \\
          &        &         &       &         &       &    1.060&  Sa97b&    1.890&  0.050&  Sa97b&      &       & RO \\
1928$+$738&   0.360&    3.339&  Fo98 &    1.110&  Fo98 &    0.520&  Go95 &    2.250&  0.480&  Go95 &     3&  Gh93 & EX \\
          &        &         &       &         &       &    1.470&  Fo98 &    2.330&  0.190&  Fo98 &      &       & RO \\
\hline
\end{tabular}
\caption{column (1): name of the source (IAU); 
(2): redshift; 
(3): radio flux at 5 GHz in Jy; 
(4): reference for the radio flux;
(5): optical flux at 5500 \AA; 
(6): reference for the optical flux; 
(7): X--ray flux at 1 keV in $\mu$Jy;
(8): reference for the X--ray flux; 
(9): X--ray photon spectral index;
(10): error on the X--ray spectral index;
(11): reference for the X--ray spectral index;
(12): class (1=HBL, 2=LBL, 3=FSRQ); 
(13): reference for class; 
(14): observing satellite (AS=ASCA, RO={\it ROSAT}, SA=BeppoSAX, EX=EXOSAT, EI=Einstein).
}
\end{table*}

\begin{table*}
\begin{tabular}{ll|ll|ll}
\hline \hline
Ba94: &Bade et al. (1994)       &Gh99: &Ghisellini et al. (1999)       &Pa95: &Padovani et al. (1995)\\
Ba98: &Bade et al. (1998)       &BSGh: &Ghisellini in BeppoSAX Archive &Pa99: &Padovani et al. (1999)\\
Br85: &Bregman et al. (1985)    &Gi98: &Giommi et al. (1998)           &Pe96: &Perlman et al. (1996)\\
Br94: &Brinkmann et al. (1994)  &Gi99: &Giommi et al. (1999)           &Sa94: &Sambruna et al. (1994)\\
Br97: &Brinkmann et al. (1997)  &Go95: &Ghosh et al. (1995)            &Sa97a:&Sambruna et al. (1997)\\
Ca97: &Cappi et al. (1997)      &Gr96: &Greiner et al. (1996)          &Sa97b:&Sambruna et al. (1997)\\
Co95: &Comastri et al. (1995)   &Kh81: &Khuer et al. (1981)            &Sa99: &Sambruna et al. (1999)\\
Co97: &Comastri et al. (1997)   &Ku98: &Kubo et al. (1998)             &Sb96: &Siebert et al. (1996)\\
Cs00: &Costamante et al. (2000) &IT90: &Impey et al. (1990)            &Sb99: &Siebert et al. (1999)\\
El99: &Elvis et al. (1999)      &IT91: &Impey et al. (1991)            &Sn97: &Sing et al. (1997)\\
Fa97: &Fabian et al. (1997)     &La96: &Lamer et al. (1996)            &Ta00: &Tagliaferri et al. (2000)\\
Fa98: &Fabian et al. (1998)     &Le97: &Leighly et al. (1997)          &Tv99: &Tavecchio et al. (2000)\\
Fo98: &Fossati et al. (1998)    &LM99: &Laurent-Muehleisen et al. (1999)&VV93:&Veron-Cetty et al. (1993)\\
Fo98b:&Fossati et al. (1998)    &Ma98b:&Maraschi et al. (1998)         &Xu99: &Xue et al. (1999)\\
Ge96: &George et al. (1996)     &BSMc: &Maccacaro in BeppoSAX Archive  &Ya98: &Yaqoob et al. (1998)\\
Gh93: &Ghisellini et al. (1993) &Na96: &Nass et al. (1996)             &Wa99: &Watson et al. (1999)\\
Gh98: &Ghisellini (1998)        &NED:  &NASA-IPAC Extragalactic Database &Wo98:&Wolter et al. (1998)\\
Gh98b:&Ghisellini et al. (1998) &Or98: &Orr et al. (1998)              &Wo90: & Worrall et al. (1990)\\
\hline
\end{tabular}
\caption{
References used in Table~\ref{tab:listone}. TW means This Work.
\label{tab:references}}
\end{table*}


\begin{thebibliography}{}

\bibitem[]{} Bade N., Fink H.H. \& Engels D., 1994, A\&A, 286, 381

\bibitem[]{} Bade N., Beckmann V., Douglas N.G. Barthel P.D., Engels D.,
      Cordis L., Nass P. \& Voges W., 1998, A\&A, 334, 459

\bibitem[]{} Bregman J.N., Glassgold A.E., Huggins P.J. \& Kinney A.L., 
          1985, ApJ, 291, 505

\bibitem[]{} Brinkmann W. \& Siebert J., 1994, A\&A, 285, 812

\bibitem[]{} Brinkmann W., Siebert J., Feigelson E.D. et al., 1997, A\&A, 323, 739

\bibitem[]{} Cappi M., Matsuoka M., Comastri A., Brinkmann W., Elvis M., 
      Palumbo G.G.C. \& Vignali C., 1997, ApJ, 478, 492

\bibitem[]{} Comastri A., Molendi S. \& Ghisellini G., 1995, MNRAS, 277, 297

\bibitem[]{} Comastri A., Fossati G., Ghisellini G. \& Molendi S., 1997, ApJ, 
      480, 534

\bibitem[]{} Costamante L., Ghisellini G., Giommi P. et al., 2000, in "X--ray 
      '99: Stellar Endpoints, AGN and the Diffuse Background", in press
       (astro--ph/0001410)

\bibitem[]{} Danly L., Lockman F. J., Meade M. R. \& Savage B. D. 1992, ApJS, 81, 125

\bibitem[]{} Dickey J.M. \& Lockman F.J., 1990, ARA\&A, 28, 215

\bibitem[]{} Elvis M., Lockman F. J.  \& Wilkes B. J. 1989, AJ, 97, 777

\bibitem[]{} Elvis M., Fiore F., Siemiginowska A., Mathur S., 
          McDowell J. \& Bechtold J., 1999, HEAD, 31, 2305

\bibitem[]{} Fabian A.C., Brandt W.N., McMahon R.G. \& Hook I.M., 1997, 
            MNRAS, 291, L5

\bibitem[]{} Fabian A.C., Iwasawa K., McMahon R.G., Celotti A., Brandt W.N. \& 
            Hook I.M., 1998, MNRAS, 295, L25

\bibitem[]{} Fossati G., Celotti A., Ghisellini G. \& Maraschi L., 1997, 
      MNRAS, 289, 136

\bibitem[]{} Fossati G., Maraschi L., Celotti A., Comastri A. \& Ghisellini G.,
       1998, MNRAS, 299, 433

\bibitem[]{} Fossati G., Chiappetti L., Celotti A. et al., 1998, 
      Nuclear Physics B (p.s.) 69/1-3, 423 

\bibitem[]{} George I.M. \& Turner T.J., 1996, ApJ, 461, 198

\bibitem[]{} Ghisellini G., Padovani P., Celotti A. \& Maraschi L., 
      1993, ApJ, 407, 65

\bibitem[]{} Ghisellini G. \& Madau P., 1996, MNRAS, 280, 67

\bibitem[]{} Ghisellini G., 1998, Nuclear Physics B (p.s.) 69/1-3, 397

\bibitem[]{} Ghisellini G., Tagliaferri G., Costamante L. et al., 1998, 
       Nuclear Physics B (p.s.) 69/1-3, 427

\bibitem[]{} Ghisellini G., Tagliaferri G. \& Giommi P., 1999, IAUC 7116

\bibitem[]{} Giommi P., Padovani P. \& Perlman E., 1998, Nuclear Physics B 
           (p.s.) 69/1-3, 407

\bibitem[]{} Giommi P., Fiore F., Guainazzi M. et al., 1998, A\&A, 333L, 5

\bibitem[]{} Giommi P., Massaro E., Chiappetti L. et al., 1999, A\&A, 351, 59

\bibitem[]{} Ghosh K.K. \& Soundararajaperumal S., 1995, ApJS, 100, 37

\bibitem[]{} Greiner J., Danner R., Bade N., Richter G.A.,
        Kroll P., \& Komossa S., 1996, A\&A, 310, 384

\bibitem[]{} Kellermann K.I., Sramek R., Schmidt M., Shaffer D.B. \&
        Green R., 1989, AJ, 98, 1195

\bibitem[]{} Kuehr H., Witzel A., Pauliny-Toth I.I.K. \& Nauber U., 1981, 
      A\&AS, 45, 367

\bibitem[]{} Kubo H., Takahashi T., Madejski G., Tashiro M.,
     Makino F., Inoue S. \& Takahara F., 1998, ApJ, 504, 693

\bibitem[]{} Impey C.D. \& Tapia S., 1990, ApJ, 354, 124

\bibitem[]{} Impey C.D., Lawrence C.R. \& Tapia S., 1991, ApJ, 375, 46

\bibitem[]{} Lamer G., Brunner H. \& Staubert R., 1996, A\&A, 311, 384

\bibitem[]{} Leighly K.M., O'Brien P.T., Edelson R., George, I.M., Malkan, M.A.,
 Matsuoka M., Mushotzky R.F. \& Peterson B.M., 1997, ApJ, 483, 767

\bibitem[]{} Laurent--Muehleisen S.A., Kollgaard R.I., Ryan P.J., Feigelson E.D., 
  Brinkmann W. \& Siebert J., 1997, A\&AS, 122, 235

\bibitem[]{} Laurent--Muehleisen S.A., Kollgaard R.I., Feigelson E.D., Brinkmann W. 
  \& Siebert J., 1999, ApJ, 525, 127 

\bibitem[]{} Lockman F. J. \& Savage B. D. 1995, ApJS, 97, 1

\bibitem[]{} Mannheim K., 1993, A\&A, 269, 67

\bibitem[]{} Maraschi L., Ghisellini G. \& Celotti A., 1992, ApJ, 397L, 5

\bibitem[]{} Maraschi L., Celotti A., Fossati G. et al., 1998, Nuclear 
             Physics B (p.s.) 69/1-3, 453

\bibitem[]{} Murphy E. M., Lockman F. J., Laor A., \& Elvis M. 1996, ApJS, 105, 36

\bibitem[]{} Nass P., Bade N., Kollgaard R.I., Laurent--Muehleisen S A.,
     Reimers D. \& Voges W., 1996, A\&A, 309, 419

\bibitem[]{} Orr A., Yaqoob T., Parmar A.N., Piro L., White N.E.  
     \& Grandi P., 1998, A\&A, 337, 685

\bibitem[]{} Padovani P. \& Giommi P., 1995, MNRAS, 277, 1477

\bibitem[]{} Padovani P., Morganti R., Siebert J., Siebert J.,
     Vagnetti F. \& Cimatti A., 1999, MNRAS, 304, 829 

\bibitem[]{} Perlman E.S., Stocke J.T., Wang Q.D. \& Morris, S.L., 
     1996, ApJ, 456, 451

\bibitem[]{} Reeves J.N., Turner M.J.L., Ohashi T. \& Kii T.,
     1997, MNRAS, 292, 468

\bibitem[]{} Sambruna R.M., Barr P., Giommi P., Maraschi L.,
    Tagliaferri G. \& Treves, A., 1994, ApJS, 95, 371

\bibitem[]{} Sambruna R.M., George I.M., Madejski G., Urry C.M., Turner T.J.,
   Weaver K.A., Maraschi L. \& Treves, A. 1997, ApJ, 483, 774

\bibitem[]{} Sambruna R.M., 1997, ApJ, 487, 536

\bibitem[]{} Sambruna R.M., Ghisellini G., Hooper E., Kollgaard R.I., 
      Pesce J.E. \& Urry C.M.,  et al., 1999, ApJ, 515, 140

\bibitem[]{} Sambruna R.M., Chou L.L. \& Urry C.M., 2000, ApJ, 533, 650

\bibitem[]{} Siebert J., Matsuoka M., Brinkmann W., Cappi M.,
   Mihara T. \& Takahashi T., 1996, A\&A, 307, 8

\bibitem[]{} Siebert J., Brinkmann W., Gliozzi M., Laurent--Muehleisen S.A. \&
    Matsuoka M., 1999, AN, 320, 315

\bibitem[]{} Sikora M., Begelman M.C. \& Rees M.J., 1994, ApJ, 421, 153

\bibitem[]{} Sing K.P., Shrader C.R. \& George I.M., 1997, ApJ, 491, 515

\bibitem[]{} Tagliaferri G., Ghisellini G., Giommi P. et al., 2000, 
      A\&A, 354, 431

\bibitem[]{} Tanaka Y. et al., 1994, PASJ, 46, L37

\bibitem[]{} Tavecchio F., Maraschi L., Ghisellini G. et al., 2000, ApJ in press
   (astro-ph/0003019)

\bibitem[]{} Veron-Cetty M.P.; Veron P., 1993, ESO Scientific Report, Garching

\bibitem[]{} Xue S.J. \& Zhang Y.H., 2000, in "X--ray '99: Stellar Endpoints, 
     AGN and the Diffuse Background" , in press (astro-ph/9911293)

\bibitem[]{} Yaqoob T., George I.M., Turner T.J., Nandra K.,
      Ptak A. \& Serlemitsos P.J., 1998, ApJ, 505, L87

\bibitem[]{} Watson D., Hanlon L., McBreen D. et al., 1999, A\&A, 345, 414

\bibitem[]{} Wolter A., Comastri A., Ghisellini G. et al., 1998, A\&A, 335, 899

\bibitem[]{} Worrall D.M. \& Wilkes B.J., 1990, ApJ, 360, 396

\end{thebibliography}
\end{document}